\RequirePackage{etex}

\documentclass[pdftex,colorlinks,notitlepage,a4paper,12pt]{article}

\usepackage{slantsc}

\usepackage[a4paper]{geometry}

\usepackage[pdftex]{graphicx}

\usepackage{epstopdf}

\usepackage[usenames,dvipsnames]{color}
\usepackage[dvipsnames]{xcolor}

\usepackage{soul}

\newlength{\mylength}
\makeatletter
\newcommand{\mycfs}[1]{%
	\normalsize
	\@defaultunits\mylength=#1pt\relax\@nnil
	\edef\@tempa{{\strip@pt\mylength}}%
	\ifx\protect\@typeset@protect
	\edef\@currsize{\noexpand\mycfs\@tempa}
	\fi
	\mylength=1.2\mylength
	\edef\@tempa{\@tempa{\strip@pt\mylength}}%
	\expandafter\fontsize\@tempa
	\selectfont
}

\usepackage[miktex]{pdftricks2}

\usepackage[utf8]{inputenc}
\usepackage[T1]{fontenc}

\usepackage[french, german, spanish, british]{babel}
\babeltags{es = spanish}
\babeltags{de = german}
\usepackage[style=american]{csquotes}

\usepackage[en-GB]{datetime2}
\DTMlangsetup[en-GB]{ord=raise}

\usepackage{textgreek}

\usepackage{latexsym}

\usepackage{amsmath}

\usepackage{econometrics}

\usepackage{amssymb}

\usepackage{pdflscape} 

\usepackage{afterpage}

\usepackage{array}
\usepackage{delarray}
\usepackage{cases}
\usepackage{rotating}
\usepackage{longtable}
\usepackage{threeparttable}
\usepackage{threeparttablex}
\usepackage{booktabs}
\usepackage{caption}
\usepackage{subcaption}
\usepackage{makecell}

\usepackage{tabularx}
\usepackage{ltablex}
\keepXColumns
\usepackage{tabu}
\usepackage{linegoal}
\usepackage{multirow}

\makeatletter
\newlength\LongtableWidth
\newcommand*{\org@longtable}{}
\let\org@longtable\longtable
\def\longtable{%
	\begingroup
	\advance\c@LT@tables\@ne
	\edef\x{LT@\romannumeral\c@LT@tables}
	\global\LongtableWidth\z@
	\@ifundefined{\x}{%
	}{%
		\def\LT@entry##1##2{%
			\global\advance\LongtableWidth##2\relax
		}%
		\@nameuse{\x}%
	}%
	\typeout{* \x: \the\LongtableWidth}%
	\endgroup \ifdim\LongtableWidth>\z@ \setlength{\LTcapwidth}{\LongtableWidth}%
	\fi
	\org@longtable }
\makeatother

\usepackage{ragged2e}

\usepackage{xpatch}

\usepackage{float}
\usepackage{placeins}

\usepackage{eurosym}    

\usepackage{appendix}

\usepackage[hyphens]{url}

\usepackage{xurl}

\hyphenchar\font=\string"7F
\usepackage[super]{nth}

\usepackage[shortlabels]{enumitem} 
\setlist[enumerate]{nosep, label = {(\arabic*)}}
\setlist[itemize]{nosep, label = {---}}

\usepackage{titlesec}
\titlelabel{\thetitle.\quad}
\titleformat*{\section}{\normalfont\large\bfseries\setstretch{1.5}}
\titleformat*{\subsection}{\normalfont\large\bfseries\setstretch{1}}

\usepackage{fancyhdr}

\makeatletter
\newread\pin@file
\newcounter{pinlineno}
\newcommand\pin@accu{}
\newcommand\pin@ext{pintmp}
\newcommand*\partialinput [3] {%
	\IfFileExists{#3}{%
		\openin\pin@file #3
		\setcounter{pinlineno}{1}
		\@whilenum\value{pinlineno}<#1 \do{%
			\read\pin@file to\pin@line
			\stepcounter{pinlineno}%
		}
		\addtocounter{pinlineno}{-1}
		\let\pin@accu\empty
		\begingroup
		\endlinechar\newlinechar
		\@whilenum\value{pinlineno}<#2 \do{%
			\readline\pin@file to\pin@line
			\edef\pin@accu{\pin@accu\pin@line}%
			\stepcounter{pinlineno}%
		}
		\closein\pin@file
		\expandafter\endgroup
		\scantokens\expandafter{\pin@accu}%
	}{%
		\errmessage{File `#3' doesn't exist!}%
	}%
}
\makeatother

\usepackage[backend=biber,style=apa,uniquelist=false,uniquename=false,dashed=true,dateabbrev=false,eventdate=comp,hyperref=true,language=british,parentracker=true]{biblatex}

\DeclareNameFormat{labelname:poss}{
	\nameparts{#1}
	\ifcase\value{uniquename}%
	\usebibmacro{name:family}{\namepartfamily}{\namepartgiven}{\namepartprefix}{\namepartsuffix}%
	\or
	\ifuseprefix
	{\usebibmacro{name:first-last}{\namepartfamily}{\namepartgiveni}{\namepartprefix}{\namepartsuffixi}}
	{\usebibmacro{name:first-last}{\namepartfamily}{\namepartgiveni}{\namepartprefixi}{\namepartsuffixi}}%
	\or
	\usebibmacro{name:first-last}{\namepartfamily}{\namepartgiven}{\namepartprefix}{\namepartsuffix}%
	\fi
	\usebibmacro{name:andothers}%
	\ifnumequal{\value{listcount}}{\value{liststop}}{'s}{}}
\DeclareFieldFormat{shorthand:poss}{%
	\ifnameundef{labelname}{#1's}{#1}}
\DeclareFieldFormat{citetitle:poss}{\mkbibemph{#1}'s}
\DeclareFieldFormat{label:poss}{#1's}
\newrobustcmd*{\posscitealias}{%
	\AtNextCite{%
		\DeclareNameAlias{labelname}{labelname:poss}%
		\DeclareFieldAlias{shorthand}{shorthand:poss}%
		\DeclareFieldAlias{citetitle}{citetitle:poss}%
		\DeclareFieldAlias{label}{label:poss}}}
\newrobustcmd*{\posscite}{%
	\posscitealias%
	\textcite}
\newrobustcmd*{\Posscite}{\bibsentence\posscite}
\newrobustcmd*{\posscites}{%
	\posscitealias%
	\textcites}

\DeclareFieldFormat*{edition}{\nth{#1} ed.}

\newcommand{\noop}[1]{}

\usepackage{siunitx} 

\def\yyy{%
	\bgroup\uccode`\~\expandafter`\string-%
	\uppercase{\egroup\edef~{\noexpand\text{\llap{\textendash}\relax}}}%
	\mathcode\expandafter`\string-"8000 }

\def\xxxl#1{%
	\bgroup\uccode`\~\expandafter`\string#1%
	\uppercase{\egroup\edef~{\noexpand\text{\noexpand\llap{\string#1}}}}%
	\mathcode\expandafter`\string#1"8000 }

\def\xxxr#1{%
	\bgroup\uccode`\~\expandafter`\string#1%
	\uppercase{\egroup\edef~{\noexpand\text{\noexpand\rlap{\string#1}}}}%
	\mathcode\expandafter`\string#1"8000 }

\usepackage{setspace}
\usepackage[bottom]{footmisc}

\usepackage[framemethod=tikz]{mdframed}
\usepackage{tikz}
\usetikzlibrary{shapes.geometric, arrows, positioning, calc, decorations.markings, shapes.misc, fit}

\usepackage{acronym}
\acrodef{sso}[SSO]{Spanish Statistical Office}
\acrodef{imf}[IMF]{International Monetary Fund}
\acrodef{iom}[IOM]{International Organization for Migration}
\acrodef{gp}[GP]{general practitioner}

\usepackage{etoolbox}
\makeatletter
\patchcmd{\maketitle}{\@makefntext}{\fakecommand}{}{}
\patchcmd{\maketitle}{\rlap}{\hbox}{}{}
\patchcmd{\@maketitle}{\@author}{\hspace*{5pt}\@author}{}{}
\makeatother

\usepackage{comment}

\usepackage{changes}

\definechangesauthor[color=Green]{NA}

\usepackage{todonotes}

\usepackage{abstract}
\usepackage{titling}

\usepackage[noblocks]{authblk}

\title{Immigrant assimilation in health care utilisation in Spain\thanks{Corresponding author: Jos{\'e}-Ignacio Ant{\'o}n, Department of Applied Economics, University of Salamanca, Campus Miguel de Unamuno, 37007 Salamanca (Spain), e-mail: \href{mailto:janton@usal.es}{\texttt{janton@usal.es}}. We thank the comments from Rodrigo Ceni, Rafael Grande, Rafael Mu{\~n}oz de Bustillo and participants at the \nth{41} Health Economics Association Conference on a first draft. We acknowledge the finacial support of the Spanish Ministry of Science and Innovation (project PID2021-123875NB-I00).}}

\pretitle{\vspace{-40pt}\centering\Large\bfseries}
	\posttitle{\par\centering}

\predate{\centering}
	\postdate{\centering\vskip -2em}

\makeatletter
\AtBeginDocument{%
	\newlength{\temp@x}%
	\newlength{\temp@y}%
	\newlength{\temp@w}%
	\newlength{\temp@h}%
	\def\my@coords#1#2#3#4{%
		\setlength{\temp@x}{#1}%
		\setlength{\temp@y}{#2}%
		\setlength{\temp@w}{#3}%
		\setlength{\temp@h}{#4}%
		\adjustlengths{}%
		\my@pdfliteral{\strip@pt\temp@x\space\strip@pt\temp@y\space\strip@pt\temp@w\space\strip@pt\temp@h\space re}}%
	\ifpdf
	\typeout{In PDF mode}%
	\def\my@pdfliteral#1{\pdfliteral page{#1}}
	\def\adjustlengths{}%
	\fi
	\ifxetex
	\def\my@pdfliteral #1{}
	\def\adjustlengths{\setlength{\temp@h}{-\temp@h}\addtolength{\temp@y}{1in}\addtolength{\temp@x}{-1in}}%
	\fi%
	\def\Hy@colorlink#1{%
		\begingroup
		\ifHy@ocgcolorlinks
		\def\Hy@ocgcolor{#1}%
		\my@pdfliteral{q}%
		\my@pdfliteral{7 Tr}
		\else
		\HyColor@UseColor#1%
		\fi
	}%
	\def\Hy@endcolorlink{%
		\ifHy@ocgcolorlinks%
		\my@pdfliteral{/OC/OCPrint BDC}%
		\my@coords{0pt}{0pt}{\pdfpagewidth}{\pdfpageheight}%
		\my@pdfliteral{F}
		\my@pdfliteral{EMC/OC/OCView BDC}%
		\begingroup%
		\expandafter\HyColor@UseColor\Hy@ocgcolor%
		\my@coords{0pt}{0pt}{\pdfpagewidth}{\pdfpageheight}%
		\my@pdfliteral{F}
		\endgroup%
		\my@pdfliteral{EMC}%
		\my@pdfliteral{0 Tr}
		\my@pdfliteral{Q}%
		\fi
		\endgroup
	}%
}
\makeatother

\DeclareLabeldate{%
	\field{date}
	\field{eventdate}
	\field{origdate}
	\field{urldate}
	\field{pubstate}
	\literal{nodate}
}

\renewbibmacro*{addendum+pubstate}{%
	\printfield{addendum}%
	\iffieldequalstr{labeldatesource}{pubstate}{}
	{\newunit\newblock\printfield{pubstate}}}

\author[$\dag$]{Zuleika~Ferre}
\affil[$\dag$]{University of the Republic}
\author[$\dag$]{Patricia~Triunfo}
\author[$\ddag$]{Jos\'e-Ignacio~Antón}
\affil[$\ddag$]{University of Salamanca (Spain) and Instituto Universitario Guti{\'e}rrez Mellado, Universidad Nacional de Educaci{\'o}n a Distancia (Spain)}

\usepackage[pdftex,pdftitle={},pdfsubject={},%
pdfauthor={},breaklinks=true]{hyperref}
\hypersetup{
	colorlinks=true,%
	citecolor=blue,%
	filecolor=black,%
	linkcolor=black,%
	urlcolor=blue}

\begin{document}
	
\date{}

\maketitle

\singlespacing	


\begin{abstract}
	\noindent Abundant evidence has tracked the labour market and health assimilation of immigrants, including static analyses of differences in how foreign-born and native-born residents consume health care services. However, we know much less about how migrants' patterns of health care usage evolve with time of residence, especially in countries providing universal or quasi-universal coverage. We investigate this process in Spain by combining all the available waves of the local health survey, which allows us to separately identify period, cohort, and assimilation effects. We find that the evidence of health assimilation is limited and solely applies to migrant females' visits to general practitioners. Nevertheless, the differential effects of ageing on health care use between foreign-born and native-born populations contributes to the convergence of utilisation patterns in most health services after 20 years in Spain. Substantial heterogeneity over time and by region of origin both suggest that studies modelling future welfare state finances would benefit from a more thorough assessment of migration.\vskip 0.5em
	
	\noindent\textbf{Keywords:} migration, health care, assimilation, Spain.\vskip 0.5em
	
	\noindent\textbf{JEL classification:} J15, I10.
\end{abstract}

\singlespacing	
	
\section{Introduction}\label{Section 1}

Migrants are common in societies of many different development levels \parencite{iom2019}. Migrants' overall success largely depends on their assimilation into the host society, including in how they access and use social services. Although immigration exerts overall beneficial effects on destination countries \parencite[][Chapter~4]{imf2020}, its impact on the sustainability of the welfare state is a subject of debate. These effects most likely depend on migrants' characteristics, the specific features of welfare states, and the time horizon used to measure service use \parencite{barrett2008,christl2022,giulietti2014,hansen2017,nannestad2007}. In turn, as long as migration increases the ethnic heterogeneity of societies, it can also shape natives' attitudes towards resource redistribution and the welfare state itself \parencite{alesina2021a, alesina2021b, dahlberg2012, facchini2016}. 

This paper explores the patterns of migrants’ assimilation into health care in Spain. In particularly, we evaluate how length of residence affects migrants’ use of this service, considered the cornerstone of the Spanish welfare state. We also explore the extent of convergence in health care utilization patterns between native and migrant populations, in terms of both assimilation and differential ageing processes.

To pursue these goals, we employ different econometric models to exploit
four Spanish health surveys (conducted from 2010 to 2020). These surveys contain information on respondents’ time of residence in the country and include more than 80,000 natives and nearly 8,000 foreign-born individuals. This setup allows us to identify cohort, period, and assimilation effects and to carry out separate analysis by gender. Our findings suggest relevant migrant cohort effects in many services, i.e., foreign-born populations use some types of health care less than natives in their first years in the country. These results are consistent with both the healthy migrant effect and the persistence of information problems and language barriers.

Assimilation—in this context, the increase in health care utilization with the number of years in the host country—is only relevant for visits to general practitioners (GPs) by female migrants. We do not observe large differences by region of origin, although the statistical power of our analyses substantially decreases when examining different groups of migrants separately. Though the effect is modest, we argue that this outcome might be due to partial health assimilation and limited socio-economic progress, given that the use of health care services in Spain is higher among individuals with high educational attainment.

Even if migrants use less health care at arrival and their assimilation is quite limited, the age patterns of their utilisation are different than among locals. As a result, after 20 years in Spain, migrants tend to exhibit similar levels of effective recourse to health services as the native population, with a few exceptions. Particularly, migrant women visit GPs more often than natives after 10 years and some cohorts of foreign-born females make higher use of emergency care.

Overall, our results suggest that the differences between comparable native and migrant populations in terms of health care utilisation are relatively minor when the latter arrive to the country, and that after some time, their behaviour becomes considerably similar. Nevertheless, we identify some services where health consumption seems higher among foreign-born women. We argue that this result might indicate a modest extra cost in terms of social spending due to migration. Further simulations assessing the costs and benefits of migration in Spain might profit from these findings.

To our knowledge, this study is the first to explore immigrant assimilation based on health care use in Spain. Whereas several papers provide evidence on how foreign-born populations exhibit better health than locals on arrival, how this gap tends to close with time of residence in the country \parencite{rivera2013, rivera2015}, and how foreign-born populations exhibit similar or lower health service utilization \parentext{see, e.g., the surveys of \textcite{llop2014}, \textcite{norredam2009} and \textcite{sarriasantamera2016}, among many others}, no previous work addresses this question linked to the dynamics of health care use. 

Whereas the literature on immigrant health \parencite{antecol2015, giuntella2016a, hamilton2011} and specially labour market assimilation \parencite{abramitzky2020, lalonde1997, bodvarsson2013} is abundant, few studies consider how patterns of immigrant health care use evolve with time of residence in the host country. Studies focused on the U.S. indicate no difference in take-up rates of Medicaid means-tested benefits between natives and migrants, whose participation rises with the time spend in the country \parencite{borjas1996}, as well as higher initial health expenditures among Latino migrants than locals, with certain evidence of convergence for Latino migrants who get American citizenship \parencite{vargas2011}. 

It is also worth summarising the findings from countries providing universal access to health care. Even though, overall, these studies suggest a lower use of non-emergency health care services by migrants (vs. natives) on arrival and a certain catch-up process, this literature is not unanimous. For instance, migrant males in Canada less frequently contact a doctor on arrival, but their levels of health care usage rise to match natives after 6--8 years in the country \parencite{mcdonald2004}. According to \textcite{wadsworth2013}, the differences in health care use between migrants and natives in the United Kingdom and Germany are not large. In the U.K., foreigners make a little more use of GP services (but not hospitals) than natives. Their change in usage with time spent in the country does not follow a systematic pattern. In the case of Germany, if anything, migrants exhibit lower rates of utilisation of health care (both general practitioners and hospital services), but their rates seem to converge with those of natives with time of residence in the country. Finally, migrants' usage of primary care emergency services in Norway exhibits substantial variation over groups and is higher than natives' usage (perhaps at the expense of less effective access to other types of health care), and it decreases with length of residence in the country \parencite{smalandgoth2012a}.\footnote{\textcite{schober2022}, who only consider asylum-seekers (a very specific subset of the foreign-born population) and evaluate them five years after arrival, find that the effect of time of residence in Austria on health expenditures is negative for refugees and not clear for economic migrants: the evolution of the results for stayers is different than for the rest of the foreign-born population in this group.}

The remainder of our paper unfolds as follows. The second section provides some theoretical and institutional background for the analysis, while the third and fourth sections describe the database and methods employed in the analysis, respectively. Section 5 presents the empirical results and discusses their implications. Finally, in Section 6 we offer some concluding remarks and pathways from future research.

\section{Background}\label{Section 2}

Like most developed countries, Spain provides virtually universal health care coverage to every resident in the country. National authorities extended this entitlement in 2000 to undocumented migrants, with the sole requirement being registration in a local population census (with no legal consequences). Nevertheless, in practical terms, information problems or fear of retaliation due to their irregular status, jointly with the hesitancy of some regional authorities (who are responsible for health care delivery) to provide foreigners without legal residence in Spain with health cards, could hamper actual access to the National Health System (NHS) among some segments of migrants. 

In September 2012, in the middle of the Great Recession, the Spanish government restricted the access of undocumented foreigners to primary care (with the exceptions of minors, pregnant women, and anyone who needed emergency care). The reform still allowed the affected groups to purchase insurance for a monthly fee of \EUR{60} (individuals below 65) and \EUR{157} (individuals above 65). Some regional governments also promptly passed legislation to protect the affected populations and provide more beneficial insurance conditions, close to those before the reform. These restrictions might have resulted in less actual access to these services and worse health outcomes \parencite{jimenez2020, juanmarti2021}. 

The change in Spain's central government in 2018, after a motion of no confidence, led to the restoration of the situation prior to the limitations on migrants' health care access. In practice, similar obstacles apparently persist because of legal loopholes \parencite{villarreal2019}. \textcite{bruquetas2020} even argue that migrants' entitlement to health care in Spain has been more of a political talking point than a subject with substantial differences between the two main political parties.

Previous literature on the theoretical reasons for expecting assimilation (or non-assimilation) in health care is scant. A first reason that increased use of health services may correspond with years living in the host country, particularly relevant in contexts without universal coverage, is the higher probability of improving health care access with longer residence in a region \parencite{antecol2015,mcdonald2004,vargas2011}. A second pertinent argument has to do with the existence of the \enquote{healthy migrant effect} and a process of negative assimilation in health, documented by recent studies focused on Spain \parencite{rivera2013,rivera2015}. Similarly, during their first years in the country, migrants might face cultural and even linguistic barriers \parencite{pena2016}, whose relevance should decrease with their time spent in the country. A related argument refers to the role of information and knowledge on the functioning of health care systems, again likely to increase with time of residence in the host state \parencite{devillanova2008}. This issue could have both a negative effect on visits to GPs on arrival and a positive one on the use of emergency care \parencite{smalandgoth2010a,smalandgoth2010b}. The existence of assimilation in this area is mostly an empirical question, since there are other factors that might act in the opposite direction. For instance, keeping in mind previous studies on the impact of income on health care demand, a non-linear but overall negative trend \parencite{anton2010}, assimilation of migrants in this domain might mitigate the the increase in health care use---see, among many others, the survey of \textcite{antecol2015}. 

\section{Data}\label{Section 3}

Our analysis makes use of the National Health Survey (NHS), waves 2011--2012 and 2017 \parencite{sso2022a}, and the European Health Interview Survey (EHIS), waves 2014 and 2020 \parencite{moh2022}, administered by the \acs{sso}. These databases are the first health surveys in Spain that include precise information on the timing of immigrants' arrival in the country. The EHIS began much later than the NHS (whose first wave corresponds to 1987), but the EHIS is designed to be fully comparable to the NHS. Consequently, local authorities have discontinued the NHS, which they carried out roughly each three years until 2014. Both sources are representative of the resident population in the country aged at least 15 years old, at the regional level. Each wave includes approximately 24,000 households and follows a three-stage stratified sampling design, since it only selects one adult person from each household---households are randomly chosen from each census section---for an interview. Apart from basic socio-demographic characteristics, the questionnaires in both types of surveys cover detailed and comparable self-reported information on health status and health care utilisation. The main differences between the two sources is that the earlier questionnaire includes additional items on quality of life (e.g., information on social support) and additionally interviews a minor living each household. Hereafter, we refer to both questionnaires as national health surveys (NHSs).

For the purpose of this investigation, we pool the samples of adults in the four waves mentioned above. We identify migrant status by looking at the country of birth rather than citizenship, because naturalization processes in Spain differ widely by state of origin (e.g., they are much shorter for people from some Latin American and Caribbean countries). Regarding health care use, we focus on the following items related to health services use: number of visits to general practitioners (GPs) in the last four weeks, number of visits to specialist doctors in the last four weeks, number of hospitalisations in the last 12 months, and number of times the person used emergency care in the last 12 months. 

The resulting sample, after dropping the observations with missing values on any of the variables included in the analysis (1.1\% of cases), comprises 80,122 native and 7,807 migrant adults. Using survey weights, the latter group represents 13.8\% of the sample. Thanks to ad hoc agreements with the Spanish institutions responsible for granting access to the data (the SSO and the Ministry of Health, respectively), we are able to distinguish among foreign-born individuals from different regions of origin. The most relevant groups in demographic terms are migrants from Latin America and the Caribbean (43.4\% of all the foreign-born adult population), those from European countries other than the European Union 15 (EU15) countries (19.7\%), those from Africa (17.5\%), and those from EU15 countries (13.6\%). In contrast to other countries like Sweden or Germany, where refugees represent an important part of of the foreign population, the motivations for immigration to Spain are overwhelmingly economic, with the exception of EU15 migrants, who are mainly attracted by the benevolent climate and the lower cost of living than their countries of origin \parencite{sso2022b}.\footnote{There is a small share of migrants from the Southern Cone of Latin America who immigrated to Spain for political reasons in the 1980s \parencite{bustillo2010}.}  

We show the main summary statistics of the sample in Table~\ref{Table 1}. It includes both the variables used in our analyses of health care use assimilation and those considered for exploring the channels through which such a process takes place (several health outcomes like self-perceived health, overweight status, and mental health problems).

\begin{singlespace}
	\begin{table}[!ht]
		\begin{ThreePartTable}
			\def\sym#1{\ifmmode^{#1}\else\(^{#1}\)\fi}
			\footnotesize
			\renewcommand{\arraystretch}{0.8}
			\begin{TableNotes}[flushleft]\setlength\labelsep{0pt}\footnotesize\justifying
				\item\textit{Note}: The number of observations is lower in the case of overweight (\num{82856}) and mental health problems (\num{87902}).	
				\item\textit{Source}: Authors' analysis from national health surveys.			
			\end{TableNotes}
			\begin{tabularx}{\textwidth}{X *{2}{S[table-column-width=2.5cm, parse-numbers=false]}}
				\caption{Descriptive statistics} \label{Table 1}\\
				\toprule
				&\multicolumn{2}{c}{\makecell{Means \\ (standard deviations)}} \\ [2ex]
				&\multicolumn{1}{c}{Natives}&\multicolumn{1}{c}{Migrants}\\
				\midrule
                No. of visits to a GP (last four weeks)&       0.361&       0.299\\
                            &     (0.767)&     (0.637)\\
                No. of visits to a specialist (last four weeks)&       0.137&       0.100\\
                            &     (0.546)&     (0.445)\\
                No. of hospitalisations (last year)&       0.131&       0.120\\
                            &     (1.438)&     (1.661)\\
                No. of visits to emergency care&       0.471&       0.509\\
                            &     (1.403)&     (1.267)\\
                Good or very good health (last year)&       0.715&       0.770\\
                            &     (0.452)&     (0.421)\\
                Overweight (body mass index $\geq$ 25)&       0.529&       0.500\\
                            &     (0.499)&     (0.500)\\
                Mental health problems (last year)&       0.118&       0.075\\
                            &     (0.322)&     (0.264)\\
                Female      &       0.505&       0.535\\
                            &     (0.500)&     (0.499)\\
                Age         &      49.480&      40.321\\
                            &    (19.173)&    (14.504)\\
                Married     &       0.575&       0.565\\
                            &     (0.494)&     (0.496)\\
                Low education&       0.619&       0.507\\
                            &     (0.486)&     (0.500)\\
                Medium education&       0.196&       0.325\\
                            &     (0.397)&     (0.468)\\
                High education&       0.185&       0.168\\
                            &     (0.389)&     (0.374)\\
                Employed    &       0.452&       0.533\\
                            &     (0.498)&     (0.499)\\
                Unemployed  &       0.116&       0.200\\
                            &     (0.320)&     (0.400)\\
                Inactive    &       0.433&       0.266\\
                            &     (0.495)&     (0.442)\\
                Arrived before 1996&            &       0.133\\
                            &            &     (0.339)\\
                Arrived between 1996 and 2007&            &       0.623\\
                            &            &     (0.485)\\
                Arrived after 2007 &            &       0.244\\
                            &            &     (0.430)\\
                Less than 5 years since migration&            &       0.131\\
                            &            &     (0.338)\\
                5--9 years since migration&            &       0.245\\
                            &            &     (0.430)\\
                10 or more years since migration&            &       0.624\\
                            &            &     (0.484)\\
                [1ex]
                No. of observations&\multicolumn{1}{c}{\hspace{2mm}\num{80122}}&\multicolumn{1}{c}{\hspace{3.5mm}\num{7807}}\\
				\bottomrule	
				\insertTableNotes
			\end{tabularx}
		\end{ThreePartTable}
	\end{table}
\end{singlespace}
\FloatBarrier

\section{Methods}\label{Section 4}

In order to disentangle the effect of foreigners' length of residence on their patterns of health care use, we adopt the empirical strategy utilized by \textcite{antecol2006, antecol2015} and \textcite{giuntella2016b} for analysing health assimilation. Specifically, we estimate equations of the following form:
\begin{equation}
	\label{eq1}
	Y_{i} = {X_i}\beta + {A_i}\delta + {C_i}\gamma + {T_i}\pi + \varepsilon_{i},
\end{equation}
where $Y_i$ denotes a health care variable relevant to person $i$, $X_i$ a vector of socio-demographic control variables (region, degree of urbanisation, and a cubic of age, fully interacted with migrant status, education, and marital and activity status.), $A_i$ a vector of dummy variables indicating how long an immigrant has lived in Spain (set equal to 0 for locals and excluding a category that serves as reference), ${C_i}$ a vector of dummy variables identifying the arrival cohort (which takes the value 0 for natives), $T_i$ a vector of dummy variables capturing the survey year, and $\varepsilon_{i}$ a random disturbance. 

We estimate equation~\ref{eq1} by Ordinary Least Squares (OLS) in our baseline analysis. We are not interested in prediction but in the marginal effects of the dummies due to cohort and assimilation variables. In this respect, OLS estimates are consistent under weaker assumptions than others obtained from count data models \parencite{angrist2008}. In any case, in Subsection~\ref{Section 5.4}, devoted to robustness checks, we present the results obtained using count data models.

As there are relevant differences in health and health care use by gender, we estimate the equation of interest separately for men and women. As in the case of health status \parencite{antecol2015,giuntella2022}, ethnicity might also play a role in health services utilisation. Whereas most of the population born in Spain is white, there is a considerable heterogeneity in the ethnic composition of migrant adults. This is due to the relevance of migration from Latin America, the Caribbean, and Africa. Previous research has identified relevant differences in the patterns of migrant health care use by state of origin \parencite{llop2014}. For this reason, we additionally re-estimate our model comparing locals with migrants from the EU15, the rest of Europe, Latin America, and the Caribbean and Africa. Unfortunately, considering these groups separately largely reduces the available samples, which makes the estimates quite imprecise.  

Pooling several cross-sections and both migrants and natives allows us to separately identify cohort, assimilation, and period effects. We consider three arrival cohorts based on Spain's recent economic and social conditions: before 1996 (when massive immigration began), 1996--2007 (a period of strong economic growth before the financial crisis), and 2008--2020 (just after the start of the Great Recession). In order to study assimilation, we take into account three intervals of length of stay in Spain: 0--4, 5--9, and 10 or more years. In order to identify the model, we omit the first category of time of residence. The coefficients due to arrival cohort indicate the differences in health care utilisation between migrants and natives at 0--4 years since migration. Those associated with the two dummies of length of stay in the country (5--9 and 10 or more years) capture the change in health care use for migrants with time spent in Spain. Combining the coefficients of the interaction between age and migrant status, the arrival cohort, and the time since arrival allow us to calculate how the migrant-native gap in health care use evolves over time. We can identify the period effect thanks to the inclusion of natives in the sample.   

In principle, the coefficients of the binary variables due to migrant cohorts would indicate the gap in health care demand between locals and migrants on arrival at 0 years old. Therefore, to make the interpretation of the results easier, we centre age at 15 years old (the lowest age at which we can observe individuals in our database), so those parameters capture the difference between foreign-born individuals at 0--4 years since migration and native population at that age.  

We explicitly refrain from introducing variables controlling for health status in the left-hand side of equation, as those sort of variables are jointly determined with utilisation of health care services, and both are part of the process of immigrants' assimilation to their host countries. In order to explore the potential role of these factors in shaping the use patterns of health services, we further estimate the role of immigrant assimilation in a set of health outcomes (self-reported health status, overweight status, and prevalence of mental health problems).

Selection of return migration represents a potential threat to identification. For instance, if foreign-born populations who go back to their country of origin have worse health (and higher demand for health care) than stayers, the estimated coefficients would be inconsistent (downward biased). Regrettably, although return migration became a very relevant phenomenon during the Great Recession and its aftermath \parencite{larramona2013, izquierdo2016}, there is little available evidence on the relationship between this phenomenon and health status for Spain. A survey of studies by \textcite{antecol2015}, focused on other countries, suggests mixed results, so the experience of other societies does not provide a clear guide here. If, as shown by \textcite{abramitzky2014} in the case of the labour market, negative self-selection of return migration were the norm, our estimates of assimilation would be a lower bound for the actual effect of the number of years spent by migrants in Spain.  

\section{Results}\label{Section 5}

\subsection{Main results}\label{Section 5.1}

We display the main results of the econometric analysis in Table~\ref{Table 2}, which shows the estimates due to visits to GPs and specialists, and \ref{Table 3}, which refers to hospital admissions and emergency care. Regarding the number of visits to GPs, we observe no difference in the frequency of use between locals and migrants on arrival in the case of males. Nevertheless, we observe that two cohorts of foreign-born women make a less intense use of health services that their native counterparts. The length of residence in Spain does not affect the utilisation of this service in the case of foreign-born men, but it increases after 10 years in Spain by 0.102 visits among migrant women. The size of this assimilation effect is not negligible: it represents nearly a quarter of the average number of visits to GPs. With regards to contacts with specialist doctors, assimilation is absent among both sexes, Nevertheless, it is worth mentioning that, on arrival, the female migrant cohort arriving between 2008 and 2020 exhibited a lower rate of utilisation of this service than natives. This effect (0.057 visits less) constitutes approximately one-third of the average number of contacts with specialists in Spain. 

Table~\ref{Table 3} shows the results for hospitalisations and emergency care. They reveal no difference between male migrants on arrival and locals for the former variable, but all the cohorts of foreign-born women exhibit substantially higher hospital admissions than comparable natives.\footnote{This result probably reflects the fact that migrants' fertility rates on arrival are higher than natives' rates (Spain has one of the lowest native fertility rates in the world), correlated with the ones observed in their country of origin and decreasing with time spent in Spain \parencite{grande2017}.} Also, on arrival, two cohorts of migrant women exhibit a lower use emergency services than female natives. We do not observe any cohort or assimilation effects.  

\begin{singlespace}
	\begin{table}[!ht]
		\begin{ThreePartTable}
			\def\sym#1{\ifmmode^{#1}\else\(^{#1}\)\fi}
			\footnotesize
			\begin{TableNotes}[flushleft]\setlength\labelsep{0pt}\footnotesize\justifying
				\item\textit{Notes}: \sym{***} significant at 1\% level; \sym{**} significant at 5\% level; \sym{*} significant at 10\% level. All specifications include an intercept, year and region fixed effects, degree of urbanisation, education and marital and activity status. Heterokedasticity-robust standard errors in parentheses.
				\item\textit{Source}: Authors' analysis from national health surveys.				
			\end{TableNotes}
			\begin{tabularx}{\textwidth}{X *{4}{S[table-column-width=2cm]}}
				\caption{Age, immigrant arrival cohort and assimilation effects (OLS) in visits to GP and specialist} \label{Table 2}\\
				\toprule
				&\multicolumn{1}{c}{(I)}&\multicolumn{1}{c}{(II)}&\multicolumn{1}{c}{(III)}&\multicolumn{1}{c}{(IV)}\\[1ex]
				&\multicolumn{2}{c}{No. of visits to GP}&\multicolumn{2}{c}{No. of visits to specialist} \\[1ex]
				&\multicolumn{1}{c}{Men}&\multicolumn{1}{c}{Women}&\multicolumn{1}{c}{Men}&\multicolumn{1}{c}{Women}\\
				\midrule
				Age effects and interactions&&& \\ [6pt]
                ~~Age       &       0.025\sym{***}&      -0.000         &       0.018\sym{***}&       0.005         \\
                            &     (0.006)         &     (0.006)         &     (0.005)         &     (0.003)         \\
                ~~$\text{Age}^2 /100$&      -0.038\sym{***}&       0.008         &      -0.029\sym{***}&      -0.004         \\
                            &     (0.012)         &     (0.013)         &     (0.010)         &     (0.006)         \\
                ~~$\text{Age}^3$ /10000&       0.024\sym{***}&      -0.004         &       0.014\sym{**} &       0.000         \\
                            &     (0.008)         &     (0.008)         &     (0.006)         &     (0.004)         \\
                ~~$\text{Age} \times \text{migrant}$&      -0.003         &       0.029\sym{**} &      -0.005         &       0.006         \\
                            &     (0.013)         &     (0.013)         &     (0.007)         &     (0.007)         \\
                ~~$\text{Age}/100 \times \text{migrant}$&       0.003         &      -0.061\sym{**} &       0.010         &      -0.014         \\
                            &     (0.031)         &     (0.031)         &     (0.016)         &     (0.016)         \\
                ~~$\text{Age}/\num{10000} \times \text{migrant}$&      -0.001         &       0.035         &      -0.006         &       0.009         \\
                            &     (0.022)         &     (0.022)         &     (0.011)         &     (0.011)         \\
				Immigrant arrival cohort&&& \\ [6pt]
				~~Pre--1996 &       0.022         &      -0.145\sym{*}  &       0.022         &      -0.086         \\
				&     (0.060)         &     (0.075)         &     (0.039)         &     (0.061)         \\
				~~1996--2007&       0.051         &      -0.091         &       0.024         &      -0.073         \\
				&     (0.054)         &     (0.062)         &     (0.035)         &     (0.052)         \\
				~~2008--2020&       0.029         &      -0.123\sym{**} &      -0.007         &      -0.067\sym{**} \\
				&     (0.046)         &     (0.055)         &     (0.030)         &     (0.033)         \\
				Time of residence in Spain&&& \\ [6pt]
                ~~5--9 years&      -0.048         &       0.018         &      -0.014         &      -0.014         \\
                            &     (0.039)         &     (0.043)         &     (0.024)         &     (0.036)         \\
                ~~10 or more years&      -0.022         &       0.102\sym{**} &      -0.008         &       0.054         \\
                            &     (0.041)         &     (0.046)         &     (0.028)         &     (0.045)         \\ [6pt]
                Adjusted R\textsuperscript{2} & 0.052 & 0.032 & 0.031& 0.033 \\
                No. of observations & \multicolumn{1}{c}{\hspace{2mm}\num{40936}} & \multicolumn{1}{c}{\hspace{2mm}\num{46993}} & \multicolumn{1}{c}{\hspace{2mm}\num{40936}} & \multicolumn{1}{c}{\hspace{2mm}\num{46993}} \\
                Mean of dependent variable & 0.293 & 0.410 & 0.107 & 0.156 \\
            	\bottomrule	
				\insertTableNotes
			\end{tabularx}
		\end{ThreePartTable}
	\end{table}
\end{singlespace}
\clearpage

\begin{singlespace}
	\begin{table}[!ht]
		\begin{ThreePartTable}
			\def\sym#1{\ifmmode^{#1}\else\(^{#1}\)\fi}
			\footnotesize
			\begin{TableNotes}[flushleft]\setlength\labelsep{0pt}\footnotesize\justifying
				\item\textit{Notes}: \sym{***} significant at 1\% level; \sym{**} significant at 5\% level; \sym{*} significant at 10\% level. All specifications include an intercept, year and region fixed effects, degree of urbanisation, education and marital and activity status. Heterokedasticity-robust standard errors in parentheses.
				\item\textit{Source}: Authors' analysis from national health surveys.				
			\end{TableNotes}
			\begin{tabularx}{\textwidth}{X *{4}{S[table-column-width=2cm]}}
				\caption{Age, immigrant arrival cohort and assimilation effects (OLS) in hospital stays and visits to emergency care} \label{Table 3}\\
				\toprule
				&\multicolumn{1}{c}{(I)}&\multicolumn{1}{c}{(II)}&\multicolumn{1}{c}{(III)}&\multicolumn{1}{c}{(IV)}\\[1ex]
				&\multicolumn{2}{c}{No. of hospitalisations}&\multicolumn{2}{c}{No. of visits to emergency care} \\[1ex]
				&\multicolumn{1}{c}{Men}&\multicolumn{1}{c}{Women}&\multicolumn{1}{c}{Men}&\multicolumn{1}{c}{Women}\\
				\midrule
				Age effects and interactions&&& \\ [6pt]
                ~~Age       &       0.043\sym{**} &       0.031         &       0.059\sym{***}&       0.032\sym{**} \\
                            &     (0.017)         &     (0.019)         &     (0.010)         &     (0.014)         \\
                ~~$\text{Age}^2 /100$&      -0.081\sym{**} &      -0.056         &      -0.130\sym{***}&      -0.087\sym{***}\\
                            &     (0.033)         &     (0.037)         &     (0.020)         &     (0.027)         \\
                ~~$\text{Age}^3$ /10000&       0.049\sym{***}&       0.032         &       0.086\sym{***}&       0.062\sym{***}\\
                            &     (0.019)         &     (0.021)         &     (0.011)         &     (0.016)         \\
                ~~$\text{Age} \times \text{migrant}$&      -0.042         &      -0.009         &      -0.030         &       0.057\sym{**} \\
                            &     (0.038)         &     (0.016)         &     (0.024)         &     (0.023)         \\
                ~~$\text{Age}/100 \times \text{migrant}$&       0.110         &       0.003         &       0.067         &      -0.115\sym{**} \\
                            &     (0.099)         &     (0.042)         &     (0.054)         &     (0.049)         \\
                ~~$\text{Age}/\num{10000} \times \text{migrant}$&      -0.080         &       0.004         &      -0.051         &       0.069\sym{**} \\
                            &     (0.072)         &     (0.031)         &     (0.037)         &     (0.032)         \\
				Immigrant arrival cohort&&& \\ [6pt]
				~~Pre--1996 &       0.173         &       0.243\sym{***}&       0.190         &      -0.302\sym{*}  \\
				&     (0.182)         &     (0.090)         &     (0.142)         &     (0.166)         \\
				~~1996--2007&       0.240         &       0.170\sym{**} &       0.115         &      -0.185         \\
				&     (0.233)         &     (0.072)         &     (0.108)         &     (0.152)         \\
				~~2008--2020&       0.230         &       0.116\sym{**} &       0.035         &      -0.240\sym{**} \\
				&     (0.261)         &     (0.053)         &     (0.092)         &     (0.122)         \\
                Time of residence in Spain&&& \\ [6pt]
                ~~5--9 years&      -0.170         &      -0.032         &      -0.001         &      -0.071         \\
                            &     (0.221)         &     (0.034)         &     (0.070)         &     (0.102)         \\
                ~~10 or more years&      -0.232         &      -0.007         &       0.050         &       0.062         \\
                            &     (0.239)         &     (0.067)         &     (0.078)         &     (0.109)         \\ [6pt]
                Adjusted R\textsuperscript{2} & 0.007 & 0.001 & 0.012& 0.014 \\
                No. of observations & \multicolumn{1}{c}{\hspace{2mm}\num{40936}} & \multicolumn{1}{c}{\hspace{2mm}\num{46993}} & \multicolumn{1}{c}{\hspace{2mm}\num{40936}} & \multicolumn{1}{c}{\hspace{2mm}\num{46993}} \\
                Mean of dependent variable & 0.118 & 0.142 & 0.403 & 0.548 \\
				\bottomrule	
				\insertTableNotes
			\end{tabularx}
		\end{ThreePartTable}
	\end{table}
\end{singlespace}
\FloatBarrier

\subsection{Heterogeneity by origin}\label{Section 5.2}

In this subsection, we explore how the results on assimilation to health care differ by country of origin. Specifically, we look at foreign-born populations in the four most relevant groups of migrants: EU15, other European countries, Latin America, and the Caribbean and Africa.\footnote{The rest of the immigrants represent less than 6\% and less than 1\% of the foreign-born and total populations, respectively. They include individuals from very different origins, like the United States or Asia. For these reasons, our analysis does not consider them.} Because of sample size limitations, in many cases the estimated coefficients are not statistically different from zero, though they are not statistically different from our main results either. Therefore, we reproduce those results in the supplementary appendix and we comment on the main salient findings here.

First, regarding EU15-born individuals (Table~\ref{Table A1} and \ref{Table A2}), the most interesting finding is the positive effect of assimilation on female health care use, which does not only apply to GPs but also to specialists. Second, the results for other European migrants (Table~\ref{Table A3} and \ref{Table A4}) are completely in line with the ones presented in Section~\ref{Section 5.1}. Third, in the case of foreign-born populations from Latin America and the Caribbean (Table~\ref{Table A5} and \ref{Table A6}), we found that time spent in Spain has a positive effect on women's visits to GPs and specialists, but a negative impact on hospitalisations. Last, with regards to Africans (Table~\ref{Table A7} and \ref{Table A8}), the length of residence does not seem to affect health care use, but the estimated coefficients are not statistically different from those reported in the total sample.  

\subsection{Health care use after 10 years in Spain}\label{Section 5.3}

As the number of years spent in Spain might impact the use of health care, and the utilisation patterns for these services might be different for migrants and natives, it is possible that the implications of migration for aggregate health spending vary over time. In this respect, our analysis suggests that on arrival, migrants are not an unusual burden in terms of health spending, since their rates of utilisation are no higher than their native counterparts. The only exception is hospitalisations in the case of women. In this subsection, based on those estimates, we assess whether this impact changes after the foreign-born respondents have spent 10 years in Spain.

\begin{figure}[!ht]
	\footnotesize
	\caption{Differences in health care use between 35-year-old migrants after 10 years in Spain by arrival cohort}
	\centering 
	\includegraphics[width=0.8\textwidth]{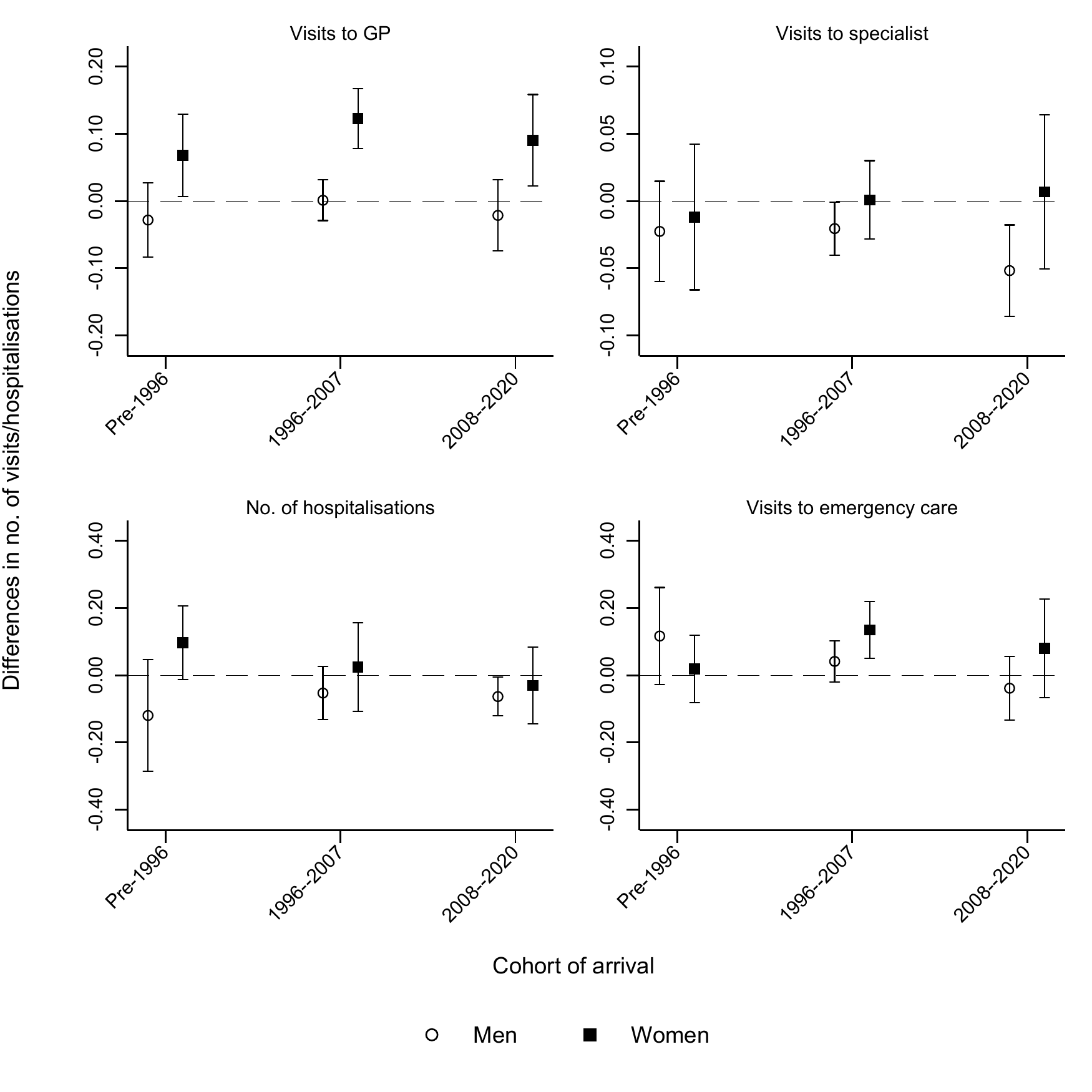} \\
	\justifying
	\noindent\textit{Note:} The graph shows point estimates and 90\%-level confidence intervals. We assume that migrants enter the country 10 years ago or earlier.\\
	\noindent\textit{Source}: Authors' analysis from results in Table~\ref{Table 2} and \ref{Table 3}.
	\label{Figure 1}
\end{figure}

Figure~\ref{Figure 1} displays the differences in health care use between migrants and locals by type of service, sex, and arrival cohort. The results indicate that, after a decade in Spain, foreign-born males do not use health care differently than natives, with the exception of specialist doctors and two arrival cohorts whose rates of utilisation are lower than natives' rates. Concerning migrant women, their number of visits to GPs is higher than natives' visits, and the same applies to the level of their use of emergency care, for migrants who arrived between 1996 and 2007. 

In the appendix (Figure~\ref{Figure A1}--\ref{Figure A4}, we present the results of separate analyses for four different groups of migrants by region of origin (EU15, other European countries, Latin America, and the Caribbean and Africa). This exercise reveals substantial heterogeneity. For instance, some cohorts of EU15 males exhibit lower rates of use of health care services than natives, in terms of visits to GPs, specialists, and emergency care, 10 years after arrival. Differences in the levels of health care utilisation between locals and foreign-born populations from other European countries are null, with the exception of the male cohort that arrived before 1996, which has more visits to specialists than comparable natives. Individuals from Latin America and the Caribbean drive the main results: after 10 years in Spain, the number of GP visits among females is higher than native women, and the same applies to two female migrant cohorts in terms of emergency care. Last, the only gaps between African migrants and locals are a higher use of primary care by the 1996---2007 female cohort, a slightly lower number of visits to specialists by males arriving before 1996, and a lower number of hospitalisations among the latest migrant cohort compared to their native counterparts.      

These results illustrate that the impact of migration on health spending can vary depending on the time horizon considered and the origin of the foreign-born population. A quick back-of-the-envelope calculation, based on the total number of uses of public health services \parencite{minsan2022} and the cost estimated by some regional health authorities for each service \parencite{dogc2013}, exemplifies this point. After 10 years of living in Spain, migrants who arrived between 1996 and 2007 required roughly 6\% higher health care spending than natives. Similarly, African individuals from the 2007--2020 arrival cohort show a lower level of consumption of these services than locals.

Another interesting outcome that emerges from this picture is that the joint effect of assimilation and differential ageing tends to neutralise the initial differential use of health services by some foreign-born cohorts. In other words, we observe a process of convergence in terms of health care utilisation between migrants and locals, which in certain cases even results in higher rates of utilisation by migrants.

\FloatBarrier

\subsection{Channels of assimilation}\label{Section 5.4}

Although the assimilation described above is somewhat limited, it is interesting to dig deeper into the potential mechanisms driving this process, even in a speculative fashion. The first possible channel is health assimilation. In this respect, using a similar econometric specification, we explore the effect of time of residence in Spain on the probability of reporting good health, overweight status, or mental health problems (Table~\ref{Table 4}). Our results suggest a pattern of negative health assimilation. In the case of females, the probability of being overweight increases by almost 8 percentage points 10 years after arrival, which is compatible with our findings revealing a higher number of visits to GPs.

The higher use of health care services among migrant women is not at odds with the findings of previous literature on Spanish immigration. Whereas the employment rates and earnings of foreign-born, working-age populations tend to increase with time of residence in Spain \parencite{amuedo2007, izquierdo2010}, their occupational assimilation is incomplete \parencite{fernandez2015,rodriguezplanas2016b,simon2014}. Bearing in mind that health care use seems to decrease with occupational attainment in Spain \parencite{lostao2011}, our results---indicating a higher number of visits to GPs by female migrants who have lived longer in Spain---would align with these labour market developments. 

In a related argument, job quality, even leaving aside pay, is substantially worse for Spanish immigrants than for locals \parencite{fernandez2007, diazserrano2013, gamero2010, galvez2022}. Recent studies document that poor working conditions might have a detrimental effect on health similar to that of unemployment \parencite{chandola2017, wang2022}. Obviously, our assessment of health assimilation is imperfect (conditioned by the survey), so job quality might be a plausible mechanism for our findings.   

A last potential channel has to do with acculturation. A relevant number of studies highlight the relevance of language and culture in migrants' health care access \parencite{fassaert2009, ndumbi2018, pena2016, smalandgoth2010a, sorensen2019, thomas2016}. Our results are in line with this literature in that the foreign-born population segment experiencing the most intense assimilation processes are Latin Americans and Caribbeans, followed by Europeans, while the number of years in Spain does not seem to affect the rates of utilisation among Africans very much. Africans are arguably the most culturally distant migrant group from Spanish locals. 


\begin{landscape}
	\vspace*{\fill}
	\begin{singlespace}
			\begin{table}[!h]
					\begin{ThreePartTable}
							\def\sym#1{\ifmmode^{#1}\else\(^{#1}\)\fi}
							\footnotesize
							\renewcommand{\arraystretch}{0.95}
							\begin{TableNotes}[flushleft]\setlength\labelsep{0pt}\footnotesize\justifying
									\item\textit{Notes}: \sym{***} significant at 1\% level; \sym{**} significant at 5\% level; \sym{*} significant at 10\% level. All specifications include an intercept, year and region fixed effects and controls for age (introduced though a third-degree polynomial fully interacted with migrant status), degree of urbanisation, education and marital and activity status. Heterokedasticity-robust standard errors in parentheses.	
									\item\textit{Source}: Authors' analysis from national health surveys.	
								\end{TableNotes}
							\begin{tabularx}{\linewidth}{X *{6}{S[table-column-width=2cm]}}
									\caption{Immigrant arrival cohort and assimilation effects (OLS) in health outcomes} \label{Table 4}\\
									\toprule
									&\multicolumn{1}{c}{(I)}&\multicolumn{1}{c}{(II)}&\multicolumn{1}{c}{(III)}&\multicolumn{1}{c}{(IV)}&\multicolumn{1}{c}{(V)}&\multicolumn{1}{c}{(VI)}\\[1ex]
									&\multicolumn{2}{c}{Good health}&\multicolumn{2}{c}{Overweight}&\multicolumn{2}{c}{\makecell{Mental health\\ problems}}\\[2ex]
									&\multicolumn{1}{c}{Men}&\multicolumn{1}{c}{Women}&\multicolumn{1}{c}{Men}&\multicolumn{1}{c}{Women}&\multicolumn{1}{c}{Men}&\multicolumn{1}{c}{Women}\\
									\midrule
									Immigrant arrival cohort&&& \\ [6pt]									
									~~Pre--1996 &      -0.056         &       0.072         &      -0.060         &      -0.086         &       0.008         &       0.028         \\
									&     (0.046)         &     (0.045)         &     (0.063)         &     (0.056)         &     (0.023)         &     (0.029)         \\
									~~1996--2007&       0.001         &       0.037         &      -0.020         &      -0.043         &      -0.019         &       0.009         \\
									&     (0.036)         &     (0.037)         &     (0.054)         &     (0.048)         &     (0.018)         &     (0.024)         \\
									~~2008--2020&       0.000         &       0.089\sym{***}&      -0.041         &       0.002         &      -0.012         &      -0.037\sym{*}  \\
									&     (0.031)         &     (0.032)         &     (0.046)         &     (0.041)         &     (0.014)         &     (0.021)         \\
									Time of residence in Spain&&& \\ [6pt]
                                    ~~5--9 years&      -0.069\sym{***}&      -0.024         &       0.049         &       0.064\sym{**} &       0.024\sym{**} &      -0.021         \\
                                                &     (0.026)         &     (0.028)         &     (0.039)         &     (0.032)         &     (0.010)         &     (0.019)         \\
                                    ~~10 or more years&      -0.077\sym{***}&      -0.033         &       0.060         &       0.078\sym{**} &       0.048\sym{***}&      -0.027         \\
                                                &     (0.027)         &     (0.030)         &     (0.041)         &     (0.036)         &     (0.013)         &     (0.021)         \\ [6pt]
                                    Adjusted R\textsuperscript{2} & 0.156 & 0.178 & 0.121& 0.134 & 0.042& 0.059 \\
                                    No. of observations & \multicolumn{1}{c}{\hspace{2mm}\num{40936}} & \multicolumn{1}{c}{\hspace{2mm}\num{46993}} & \multicolumn{1}{c}{\hspace{2mm}\num{39508}} & \multicolumn{1}{c}{\hspace{2mm}\num{43348}} & \multicolumn{1}{c}{\hspace{2mm}\num{40926}} & \multicolumn{1}{c}{\hspace{2mm}\num{46976}}\\
                                    Mean of dependent variable & 0.765 & 0.681 & 0.605 & 0.445 & 0.073 & 0.150 \\
									\bottomrule	
									\insertTableNotes
								\end{tabularx}
						\end{ThreePartTable}
				\end{table}
		\end{singlespace}
	\vfill
\end{landscape}
\FloatBarrier

\subsection{Robustness checks}\label{Section 5.5}

In this section, we comment on the results of several robustness checks that test the sensitivity of our results to different methodological choices. First, we re-estimate all our models using a Poisson regression model (Table~\ref{Table A9} and \ref{Table A10} in the online appendix). Like OLS, this approach yields consistent estimates without requiring any further function of the error term \parencite{blackburn2014, wooldridge2010}. \footnote{This model is not efficient in cases of over-dispersion. Nevertheless, other alternatives like the negative binomial regression model, which is more efficient if the error term follows the assumed functional form, result in biased and inconsistent estimates in case of the violation of such properties.} Reassuringly, the results are qualitative and qualitatively similar to our baseline estimations (Table~\ref{Table A11} and \ref{Table A12}.\footnote{One should bear in mind that, roughly, the coefficients of the Poisson regression model indicate the percentage change in the left-hand-side variable associated with a one-unit change in the right-hand-side variable.} 

Our second robustness exercise consists of performing our estimations while including only those individuals who solely have public health insurance, as a way of isolating our results from the eventual distortions due to the different normative changes in the last decade. Again, our results hold, even though the degree of precision diminishes because of the smaller sample. \footnote{Although insurance is an endogenous variable, the share of the population with only this coverage, almost all migrants and more than 60\% of locals, is representative enough to be interesting on its own.} The last sensitivity check explores whether our results vary when we limit our analysis to individuals aged less than 65 years old (Table~\ref{Table A13} and \ref{Table A14}). This methodological choice, which might help to ameliorate the eventual bias due to return migration, does not seem to have any influence on our results beyond reducing the native sample in a non-trivial way.

\section{Conclusion}\label{Section 6}

Immigration's relationship with the welfare state demands substantial attention from both policy makers and society as a whole. The access and use of health care by foreign-born populations represents a matter of relevance because of their effect on public finances, but also in terms of ensuring an adequate integration of immigrants in the host country. The results of the assessment of both issues might differ on migrants' arrival and after a longer term of residence.

This research contributes to the literature by providing detailed evidence on the process of  assimilation in health care utilisation by immigrants in Spain. Our findings suggest, first, that some population segments of foreign-born populations, on arrival, use part of these services less than comparable natives, which is compatible with the healthy migrant effect. Furthermore, such use increases with the time living in Spain, but only in the case of female migrants and visits to GPs. Although this suggests that the evidence of assimilation is not very strong, it is not negligible either: public health spending represented more than 30\% of total expenditures in this area in 2020 \parencite{dogc2013, minsan2022}. As a result of this limited assimilation and the different impact of age on the rates of health care utilisation, the patterns of consumption for these services by migrants converge with the natives' patterns and even surpass them for some female migrant cohorts and services. Note that the impact of the time spent in the host country (assimilation) and the differential effect of age on the variable of interest represent distinct phenomena. We can separate them in our analysis, thanks to the use of several waves of the database and the consideration of both natives and migrants in the main specification.    

Our results suggest that, even if the gap between migrants and natives in health care use is narrow, assimilation plays a non-negligible role. The existence of substantial heterogeneity by migrant group and time passed since arrival might influence the estimates of migrants' impact on public finances. As a consequence, we believe that researchers aiming to judge welfare state sustainability and those that cover migration could benefit from a more detailed modelling of the patterns of access to social services by foreign-born populations. 

\clearpage
\singlespacing
\printbibliography

\clearpage
\appendix
\section*{Supplementary appendix}\label{Appendix}
\setcounter{table}{0}
\setcounter{figure}{0}
\renewcommand\thetable{A.\arabic{table}}
\renewcommand\thefigure{A.\arabic{figure}}	



\begin{singlespace}
	\begin{table}[!ht]
		\begin{ThreePartTable}
			\def\sym#1{\ifmmode^{#1}\else\(^{#1}\)\fi}
			\footnotesize
			\begin{TableNotes}[flushleft]\setlength\labelsep{0pt}\footnotesize\justifying
				\item\textit{Notes}: \sym{***} significant at 1\% level; \sym{**} significant at 5\% level; \sym{*} significant at 10\% level. All specifications include an intercept, year and region fixed effects, degree of urbanisation, education and marital and activity status. Heterokedasticity-robust standard errors in parentheses.
				\item\textit{Source}: Authors' analysis from national health surveys.				
			\end{TableNotes}
			\begin{tabularx}{\textwidth}{X *{4}{S[table-column-width=2cm]}}
				\caption{Age, immigrant arrival cohort and assimilation effects (OLS) in visits to GP and specialist (foreign-born population from EU15 countries)} \label{Table A1}\\
				\toprule
				&\multicolumn{1}{c}{(I)}&\multicolumn{1}{c}{(II)}&\multicolumn{1}{c}{(III)}&\multicolumn{1}{c}{(IV)}\\[1ex]
				&\multicolumn{2}{c}{No. of visits to GP}&\multicolumn{2}{c}{No. of visits to specialist} \\[1ex]
				&\multicolumn{1}{c}{Men}&\multicolumn{1}{c}{Women}&\multicolumn{1}{c}{Men}&\multicolumn{1}{c}{Women}\\
				\midrule
				Age effects and interactions&&& \\ [6pt]
                ~~Age       &       0.027\sym{***}&      -0.000         &       0.020\sym{***}&       0.004         \\
                            &     (0.006)         &     (0.006)         &     (0.006)         &     (0.003)         \\
                ~~$\text{Age}^2 /100$&      -0.043\sym{***}&       0.009         &      -0.033\sym{***}&      -0.004         \\
                            &     (0.012)         &     (0.013)         &     (0.011)         &     (0.006)         \\
                ~~$\text{Age}^3$ /10000&       0.026\sym{***}&      -0.004         &       0.016\sym{***}&       0.000         \\
                            &     (0.008)         &     (0.008)         &     (0.006)         &     (0.004)         \\
                ~~$\text{Age} \times \text{migrant}$&       0.037         &       0.034         &      -0.014         &      -0.040\sym{**} \\
                            &     (0.023)         &     (0.038)         &     (0.012)         &     (0.019)         \\
                ~~$\text{Age}/100 \times \text{migrant}$&      -0.084         &      -0.079         &       0.021         &       0.073\sym{*}  \\
                            &     (0.052)         &     (0.081)         &     (0.025)         &     (0.038)         \\
                ~~$\text{Age}/\num{10000} \times \text{migrant}$&       0.056         &       0.049         &      -0.011         &      -0.042\sym{*}  \\
                            &     (0.036)         &     (0.054)         &     (0.017)         &     (0.023)         \\
				Immigrant arrival cohort&&& \\ [6pt]
				~~Pre--1996 &      -0.017         &      -0.313         &       0.151         &       0.093         \\
				&     (0.100)         &     (0.258)         &     (0.122)         &     (0.117)         \\
				~~1996--2007&      -0.086         &      -0.308         &       0.107         &       0.064         \\
				&     (0.091)         &     (0.214)         &     (0.120)         &     (0.108)         \\
				~~2008--2020&      -0.152         &      -0.199         &       0.074         &       0.151         \\
				&     (0.096)         &     (0.140)         &     (0.081)         &     (0.107)         \\
                Time of residence in Spain&&& \\ [6pt]
                ~~5--9 years&      -0.071         &       0.079         &      -0.016         &      -0.026         \\
                            &     (0.083)         &     (0.108)         &     (0.067)         &     (0.053)         \\
                ~~10 or more years&      -0.109         &       0.289\sym{*}  &      -0.044         &       0.116\sym{*}  \\
                            &     (0.090)         &     (0.174)         &     (0.079)         &     (0.070)         \\ [6pt]
                Adjusted R\textsuperscript{2} & 0.053 & 0.034 & 0.033& 0.033 \\
                No. of observations & \multicolumn{1}{c}{\hspace{2mm}\num{37967}} & \multicolumn{1}{c}{\hspace{2mm}\num{43279}} & \multicolumn{1}{c}{\hspace{2mm}\num{37967}} & \multicolumn{1}{c}{\hspace{2mm}\num{43279}} \\
                Mean of dependent variable & 0.302 & 0.415 & 0.112 & 0.160 \\
				\bottomrule	
				\insertTableNotes
			\end{tabularx}
		\end{ThreePartTable}
	\end{table}
\end{singlespace}
\clearpage


\vspace*{\fill}
\begin{singlespace}
	\begin{table}[!h]
		\begin{ThreePartTable}
			\def\sym#1{\ifmmode^{#1}\else\(^{#1}\)\fi}
			\footnotesize
			\begin{TableNotes}[flushleft]\setlength\labelsep{0pt}\footnotesize\justifying
				\item\textit{Notes}: \sym{***} significant at 1\% level; \sym{**} significant at 5\% level; \sym{*} significant at 10\% level. All specifications include an intercept, year and region fixed effects, degree of urbanisation, education and marital and activity status. Heterokedasticity-robust standard errors in parentheses.
				\item\textit{Source}: Authors' analysis from national health surveys.				
			\end{TableNotes}
			\begin{tabularx}{\textwidth}{X *{4}{S[table-column-width=2cm]}}
				\caption{Age, immigrant arrival cohort and assimilation effects (OLS) in hospital stays and visits to emergency care (foreign-born population from EU15 countries)} \label{Table A2}\\
				\toprule
				&\multicolumn{1}{c}{(I)}&\multicolumn{1}{c}{(II)}&\multicolumn{1}{c}{(III)}&\multicolumn{1}{c}{(IV)}\\[1ex]
				&\multicolumn{2}{c}{No. of hospitalisations}&\multicolumn{2}{c}{No. of visits to emergency care} \\[1ex]
				&\multicolumn{1}{c}{Men}&\multicolumn{1}{c}{Women}&\multicolumn{1}{c}{Men}&\multicolumn{1}{c}{Women}\\
				\midrule
				Age effects and interactions&&& \\ [6pt]
                ~~Age       &       0.044\sym{**} &       0.026         &       0.062\sym{***}&       0.030\sym{**} \\
                            &     (0.018)         &     (0.022)         &     (0.010)         &     (0.014)         \\
                ~~$\text{Age}^2 /100$&      -0.083\sym{**} &      -0.047         &      -0.138\sym{***}&      -0.084\sym{***}\\
                            &     (0.034)         &     (0.041)         &     (0.020)         &     (0.028)         \\
                ~~$\text{Age}^3$ /10000&       0.050\sym{***}&       0.027         &       0.090\sym{***}&       0.061\sym{***}\\
                            &     (0.019)         &     (0.023)         &     (0.012)         &     (0.016)         \\
                ~~$\text{Age} \times \text{migrant}$&      -0.144         &      -0.038\sym{*}  &      -0.039         &      -0.021         \\
                            &     (0.126)         &     (0.022)         &     (0.058)         &     (0.049)         \\
                ~~$\text{Age}/100 \times \text{migrant}$&       0.400         &       0.063         &       0.101         &       0.018         \\
                            &     (0.347)         &     (0.046)         &     (0.113)         &     (0.096)         \\
                ~~$\text{Age}/\num{10000} \times \text{migrant}$&      -0.300         &      -0.033         &      -0.081         &      -0.002         \\
                            &     (0.258)         &     (0.029)         &     (0.070)         &     (0.060)         \\
				Immigrant arrival cohort&&& \\ [6pt]
				~~Pre--1996 &       0.873         &       0.286\sym{**} &       0.142         &       0.026         \\
				&     (0.738)         &     (0.141)         &     (0.329)         &     (0.340)         \\
				~~1996--2007&       0.889         &       0.224\sym{*}  &      -0.109         &      -0.093         \\
				&     (0.797)         &     (0.128)         &     (0.332)         &     (0.334)         \\
				~~2008--2020&       0.922         &       0.167         &      -0.152         &       0.036         \\
				&     (0.994)         &     (0.108)         &     (0.296)         &     (0.291)         \\
				Time of residence in Spain&&& \\ [6pt]
                ~~5--9 years&      -1.027         &       0.038         &       0.132         &       0.134         \\
                            &     (1.138)         &     (0.059)         &     (0.143)         &     (0.155)         \\
                ~~10 or more years&      -1.194         &      -0.008         &       0.027         &       0.215         \\
                            &     (1.067)         &     (0.064)         &     (0.125)         &     (0.158)         \\ [6pt]
                Adjusted R\textsuperscript{2} & 0.010 & 0.001 & 0.013& 0.016 \\
                No. of observations & \multicolumn{1}{c}{\hspace{2mm}\num{37967}} & \multicolumn{1}{c}{\hspace{2mm}\num{43279}} & \multicolumn{1}{c}{\hspace{2mm}\num{37967}} & \multicolumn{1}{c}{\hspace{2mm}\num{43279}} \\
                Mean of dependent variable & 0.123 & 0.141 & 0.397 & 0.538 \\
				\bottomrule	
				\insertTableNotes
			\end{tabularx}
		\end{ThreePartTable}
	\end{table}
\end{singlespace}
\vfill
\clearpage


\vspace*{\fill}
\begin{singlespace}
	\begin{table}[!ht]
		\begin{ThreePartTable}
			\def\sym#1{\ifmmode^{#1}\else\(^{#1}\)\fi}
			\footnotesize
			\begin{TableNotes}[flushleft]\setlength\labelsep{0pt}\footnotesize\justifying
				\item\textit{Notes}: \sym{***} significant at 1\% level; \sym{**} significant at 5\% level; \sym{*} significant at 10\% level. All specifications include an intercept, year and region fixed effects, degree of urbanisation, education and marital and activity status. Heterokedasticity-robust standard errors in parentheses.
				\item\textit{Source}: Authors' analysis from national health surveys.				
			\end{TableNotes}
			\begin{tabularx}{\textwidth}{X *{4}{S[table-column-width=2cm]}}
				\caption{Age, immigrant arrival cohort and assimilation effects (OLS) in visits to GP and specialist (foreign-born population from European countries other than the EU15)} \label{Table A3}\\
				\toprule
				&\multicolumn{1}{c}{(I)}&\multicolumn{1}{c}{(II)}&\multicolumn{1}{c}{(III)}&\multicolumn{1}{c}{(IV)}\\[1ex]
				&\multicolumn{2}{c}{No. of visits to GP}&\multicolumn{2}{c}{No. of visits to specialist} \\[1ex]
				&\multicolumn{1}{c}{Men}&\multicolumn{1}{c}{Women}&\multicolumn{1}{c}{Men}&\multicolumn{1}{c}{Women}\\
				\midrule
				Age effects and interactions&&& \\ [6pt]
				~~Age       &       0.027\sym{***}&      -0.001         &       0.020\sym{***}&       0.004         \\
				&     (0.006)         &     (0.006)         &     (0.006)         &     (0.003)         \\
				~~$\text{Age}^2 /100$&      -0.043\sym{***}&       0.010         &      -0.033\sym{***}&      -0.003         \\
				&     (0.012)         &     (0.013)         &     (0.011)         &     (0.006)         \\
				~~$\text{Age}^3$ /10000&       0.026\sym{***}&      -0.005         &       0.016\sym{***}&      -0.000         \\
				&     (0.008)         &     (0.008)         &     (0.006)         &     (0.004)         \\
				~~$\text{Age} \times \text{migrant}$&       0.038         &       0.011         &      -0.019         &       0.001         \\
				&     (0.032)         &     (0.031)         &     (0.014)         &     (0.024)         \\
				~~$\text{Age}/100 \times \text{migrant}$&      -0.112         &      -0.026         &       0.035         &      -0.008         \\
				&     (0.079)         &     (0.079)         &     (0.028)         &     (0.061)         \\
				~~$\text{Age}/\num{10000} \times \text{migrant}$&       0.095         &       0.023         &      -0.025         &       0.011         \\
				&     (0.061)         &     (0.064)         &     (0.018)         &     (0.050)         \\
				Immigrant arrival cohort&&& \\ [6pt]
				~~Pre--1996 &       0.061         &      -0.221         &       0.188\sym{*}  &       0.153         \\
				&     (0.164)         &     (0.167)         &     (0.106)         &     (0.291)         \\
				~~1996--2007&      -0.036         &      -0.227\sym{**} &       0.019         &      -0.121         \\
				&     (0.115)         &     (0.103)         &     (0.066)         &     (0.106)         \\
				~~2008--2020&      -0.040         &      -0.252\sym{***}&       0.041         &      -0.073         \\
				&     (0.106)         &     (0.097)         &     (0.059)         &     (0.083)         \\
				Time of residence in Spain&&& \\ [6pt]
				~~5--9 years&      -0.041         &       0.196\sym{***}&       0.049         &       0.035         \\
				&     (0.083)         &     (0.073)         &     (0.032)         &     (0.051)         \\
				~~10 or more years&       0.036         &       0.126         &       0.107\sym{***}&       0.127         \\
				&     (0.081)         &     (0.078)         &     (0.035)         &     (0.079)         \\
                Adjusted R\textsuperscript{2} & 0.052 & 0.034 & 0.033& 0.033 \\
                No. of observations & \multicolumn{1}{c}{\hspace{2mm}\num{38065}} & \multicolumn{1}{c}{\hspace{2mm}\num{43480}} & \multicolumn{1}{c}{\hspace{2mm}\num{38065}} & \multicolumn{1}{c}{\hspace{2mm}\num{43480}} \\
                Mean of dependent variable & 0.301 & 0.413 & 0.111 & 0.159 \\
                \bottomrule	
				\insertTableNotes
			\end{tabularx}
		\end{ThreePartTable}
	\end{table}
\end{singlespace}
\vfill
\clearpage


\vspace*{\fill}
\begin{singlespace}
	\begin{table}[!h]
		\begin{ThreePartTable}
			\def\sym#1{\ifmmode^{#1}\else\(^{#1}\)\fi}
			\footnotesize
			\begin{TableNotes}[flushleft]\setlength\labelsep{0pt}\footnotesize\justifying
				\item\textit{Notes}: \sym{***} significant at 1\% level; \sym{**} significant at 5\% level; \sym{*} significant at 10\% level. All specifications include an intercept, year and region fixed effects, degree of urbanisation, education and marital and activity status. Heterokedasticity-robust standard errors in parentheses.
				\item\textit{Source}: Authors' analysis from national health surveys.				
			\end{TableNotes}
			\begin{tabularx}{\textwidth}{X *{4}{S[table-column-width=2cm]}}
				\caption{Age, immigrant arrival cohort and assimilation effects (OLS) in hospital stays and visits to emergency care (foreign-born population from European countries other than the EU15)} \label{Table A4}\\
				\toprule
				&\multicolumn{1}{c}{(I)}&\multicolumn{1}{c}{(II)}&\multicolumn{1}{c}{(III)}&\multicolumn{1}{c}{(IV)}\\[1ex]
				&\multicolumn{2}{c}{No. of hospitalisations}&\multicolumn{2}{c}{No. of visits to emergency care} \\[1ex]
				&\multicolumn{1}{c}{Men}&\multicolumn{1}{c}{Women}&\multicolumn{1}{c}{Men}&\multicolumn{1}{c}{Women}\\
				\midrule
				~~Age       &       0.040\sym{**} &       0.028         &       0.062\sym{***}&       0.034\sym{**} \\
				&     (0.018)         &     (0.021)         &     (0.010)         &     (0.014)         \\
				~~$\text{Age}^2 /100$&      -0.077\sym{**} &      -0.050         &      -0.138\sym{***}&      -0.092\sym{***}\\
				&     (0.033)         &     (0.041)         &     (0.020)         &     (0.028)         \\
				~~$\text{Age}^3$ /10000&       0.047\sym{**} &       0.029         &       0.090\sym{***}&       0.065\sym{***}\\
				&     (0.019)         &     (0.023)         &     (0.012)         &     (0.016)         \\
				~~$\text{Age} \times \text{migrant}$&       0.013         &       0.011         &       0.025         &       0.090         \\
				&     (0.019)         &     (0.022)         &     (0.051)         &     (0.060)         \\
				~~$\text{Age}/100 \times \text{migrant}$&      -0.033         &      -0.040         &      -0.049         &      -0.216         \\
				&     (0.052)         &     (0.057)         &     (0.116)         &     (0.149)         \\
				~~$\text{Age}/\num{10000} \times \text{migrant}$&       0.025         &       0.038         &       0.022         &       0.160         \\
				&     (0.045)         &     (0.045)         &     (0.081)         &     (0.115)         \\
			    Immigrant arrival cohort&&& \\ [6pt]
				~~Pre--1996 &      -0.106         &      -0.019         &       0.262         &      -0.946\sym{***}\\
				&     (0.091)         &     (0.107)         &     (0.350)         &     (0.345)         \\
				~~1996--2007&      -0.057         &      -0.063         &      -0.077         &      -0.419         \\
				&     (0.046)         &     (0.058)         &     (0.193)         &     (0.323)         \\
				~~2008--2020&      -0.054         &      -0.048         &      -0.240         &      -0.547\sym{***}\\
				&     (0.040)         &     (0.048)         &     (0.164)         &     (0.207)         \\
				Time of residence in Spain&&& \\ [6pt]
				~~5--9 years&       0.008         &       0.081         &       0.013         &       0.131         \\
				&     (0.034)         &     (0.056)         &     (0.116)         &     (0.264)         \\
				~~10 or more years&       0.029         &       0.075         &       0.083         &       0.376         \\
				&     (0.039)         &     (0.054)         &     (0.127)         &     (0.244)         \\ [6pt]   
                Adjusted R\textsuperscript{2} & 0.008 & 0.001 & 0.013& 0.015 \\
                No. of observations & \multicolumn{1}{c}{\hspace{2mm}\num{38065}} & \multicolumn{1}{c}{\hspace{2mm}\num{43480}} & \multicolumn{1}{c}{\hspace{2mm}\num{38065}} & \multicolumn{1}{c}{\hspace{2mm}\num{43480}} \\
                Mean of dependent variable & 0.118 & 0.140 & 0.400 & 0.542 \\
				\bottomrule	
				\insertTableNotes
			\end{tabularx}
		\end{ThreePartTable}
	\end{table}
\end{singlespace}
\vfill
\clearpage


\vspace*{\fill}
\begin{singlespace}
	\begin{table}[!ht]
		\begin{ThreePartTable}
			\def\sym#1{\ifmmode^{#1}\else\(^{#1}\)\fi}
			\footnotesize
			\begin{TableNotes}[flushleft]\setlength\labelsep{0pt}\footnotesize\justifying
				\item\textit{Notes}: \sym{***} significant at 1\% level; \sym{**} significant at 5\% level; \sym{*} significant at 10\% level. All specifications include an intercept, year and region fixed effects, degree of urbanisation, education and marital and activity status. Heterokedasticity-robust standard errors in parentheses.
				\item\textit{Source}: Authors' analysis from national health surveys.				
			\end{TableNotes}
			\begin{tabularx}{\textwidth}{X *{4}{S[table-column-width=2cm]}}
				\caption{Age, immigrant arrival cohort and assimilation effects (OLS) in visits to GP and specialist (foreign-born population from Latin America and the Caribbean)} \label{Table A5}\\
				\toprule
				&\multicolumn{1}{c}{(I)}&\multicolumn{1}{c}{(II)}&\multicolumn{1}{c}{(III)}&\multicolumn{1}{c}{(IV)}\\[1ex]
				&\multicolumn{2}{c}{No. of visits to GP}&\multicolumn{2}{c}{No. of visits to specialist} \\[1ex]
				&\multicolumn{1}{c}{Men}&\multicolumn{1}{c}{Women}&\multicolumn{1}{c}{Men}&\multicolumn{1}{c}{Women}\\
				\midrule
				Age effects and interactions&&& \\ [6pt]
                ~~Age       &       0.025\sym{***}&       0.002         &       0.020\sym{***}&       0.006\sym{*}  \\
                            &     (0.006)         &     (0.006)         &     (0.006)         &     (0.003)         \\
                ~~$\text{Age}^2 /100$&      -0.040\sym{***}&       0.005         &      -0.032\sym{***}&      -0.006         \\
                            &     (0.012)         &     (0.013)         &     (0.010)         &     (0.006)         \\
                ~~$\text{Age}^3$ /10000&       0.024\sym{***}&      -0.002         &       0.016\sym{***}&       0.001         \\
                            &     (0.008)         &     (0.008)         &     (0.006)         &     (0.004)         \\
                ~~$\text{Age} \times \text{migrant}$&      -0.022         &       0.059\sym{***}&      -0.004         &       0.013         \\
                            &     (0.022)         &     (0.021)         &     (0.015)         &     (0.011)         \\
                ~~$\text{Age}/100 \times \text{migrant}$&       0.053         &      -0.120\sym{**} &       0.014         &      -0.024         \\
                            &     (0.053)         &     (0.049)         &     (0.037)         &     (0.025)         \\
                ~~$\text{Age}/\num{10000} \times \text{migrant}$&      -0.037         &       0.070\sym{**} &      -0.010         &       0.013         \\
                            &     (0.039)         &     (0.035)         &     (0.028)         &     (0.017)         \\
				Immigrant arrival cohort&&& \\ [6pt]
				~~Pre--1996 &       0.065         &      -0.222\sym{**} &       0.016         &      -0.184\sym{***}\\
				&     (0.104)         &     (0.110)         &     (0.070)         &     (0.065)         \\
				~~1996--2007&       0.137         &      -0.204\sym{**} &       0.035         &      -0.128\sym{***}\\
				&     (0.093)         &     (0.089)         &     (0.057)         &     (0.048)         \\
				~~2008--2020&       0.122\sym{*}  &      -0.179\sym{**} &      -0.037         &      -0.114\sym{***}\\
				&     (0.071)         &     (0.077)         &     (0.054)         &     (0.039)         \\
				Time of residence in Spain&&& \\ [6pt]
                ~~5--9 years&      -0.095         &       0.035         &      -0.031         &       0.033         \\
                            &     (0.068)         &     (0.067)         &     (0.047)         &     (0.033)         \\
                ~~10 or more years&      -0.097         &       0.131\sym{*}  &      -0.077         &       0.080\sym{**} \\
                            &     (0.074)         &     (0.067)         &     (0.051)         &     (0.039)         \\ [6pt]
                Adjusted R\textsuperscript{2} & 0.052 & 0.033 & 0.032& 0.034 \\
                No. of observations & \multicolumn{1}{c}{\hspace{2mm}\num{38772}} & \multicolumn{1}{c}{\hspace{2mm}\num{44790}} & \multicolumn{1}{c}{\hspace{2mm}\num{38772}} & \multicolumn{1}{c}{\hspace{2mm}\num{44790}} \\
                Mean of dependent variable & 0.300 & 0.416 & 0.112 & 0.160 \\
				\bottomrule	
				\insertTableNotes
			\end{tabularx}
		\end{ThreePartTable}
	\end{table}
\end{singlespace}
\vfill
\clearpage


\vspace*{\fill}
\begin{singlespace}
	\begin{table}[!h]
		\begin{ThreePartTable}
			\def\sym#1{\ifmmode^{#1}\else\(^{#1}\)\fi}
			\footnotesize
			\begin{TableNotes}[flushleft]\setlength\labelsep{0pt}\footnotesize\justifying
				\item\textit{Notes}: \sym{***} significant at 1\% level; \sym{**} significant at 5\% level; \sym{*} significant at 10\% level. All specifications include an intercept, year and region fixed effects, degree of urbanisation, education and marital and activity status. Heterokedasticity-robust standard errors in parentheses.
				\item\textit{Source}: Authors' analysis from national health surveys.				
			\end{TableNotes}
			\begin{tabularx}{\textwidth}{X *{4}{S[table-column-width=2cm]}}
				\caption{Age, immigrant arrival cohort and assimilation effects (OLS) in hospital stays and visits to emergency care (foreign-born population from Latin America and the Caribbean)} \label{Table A6}\\
				\toprule
				&\multicolumn{1}{c}{(I)}&\multicolumn{1}{c}{(II)}&\multicolumn{1}{c}{(III)}&\multicolumn{1}{c}{(IV)}\\[1ex]
				&\multicolumn{2}{c}{No. of hospitalisations}&\multicolumn{2}{c}{No. of visits to emergency care} \\[1ex]
				&\multicolumn{1}{c}{Men}&\multicolumn{1}{c}{Women}&\multicolumn{1}{c}{Men}&\multicolumn{1}{c}{Women}\\
				\midrule
				Age effects and interactions&&& \\ [6pt]
                ~~Age       &       0.038\sym{**} &       0.028         &       0.059\sym{***}&       0.031\sym{**} \\
                            &     (0.017)         &     (0.021)         &     (0.010)         &     (0.014)         \\
                ~~$\text{Age}^2 /100$&      -0.073\sym{**} &      -0.050         &      -0.133\sym{***}&      -0.086\sym{***}\\
                            &     (0.033)         &     (0.040)         &     (0.020)         &     (0.027)         \\
                ~~$\text{Age}^3$ /10000&       0.046\sym{**} &       0.029         &       0.087\sym{***}&       0.062\sym{***}\\
                            &     (0.019)         &     (0.022)         &     (0.012)         &     (0.016)         \\
                ~~$\text{Age} \times \text{migrant}$&      -0.000         &      -0.016         &       0.002         &       0.048         \\
                            &     (0.014)         &     (0.014)         &     (0.036)         &     (0.034)         \\
                ~~$\text{Age}/100 \times \text{migrant}$&       0.005         &       0.031         &      -0.003         &      -0.090         \\
                            &     (0.034)         &     (0.030)         &     (0.086)         &     (0.074)         \\
                ~~$\text{Age}/\num{10000} \times \text{migrant}$&      -0.007         &      -0.021         &       0.002         &       0.047         \\
                            &     (0.025)         &     (0.020)         &     (0.066)         &     (0.050)         \\
				Immigrant arrival cohort&&& \\ [6pt]
				~~Pre--1996 &      -0.072         &       0.313\sym{***}&      -0.070         &      -0.040         \\
				&     (0.058)         &     (0.110)         &     (0.186)         &     (0.265)         \\
				~~1996--2007&      -0.045         &       0.162\sym{**} &       0.073         &      -0.120         \\
				&     (0.042)         &     (0.073)         &     (0.174)         &     (0.232)         \\
				~~2008--2020&      -0.037         &       0.129\sym{**} &       0.051         &      -0.107         \\
				&     (0.035)         &     (0.064)         &     (0.125)         &     (0.189)         \\
				Time of residence in Spain&&& \\ [6pt]
                ~~5--9 years&       0.081\sym{**} &      -0.081\sym{*}  &      -0.094         &      -0.085         \\
                            &     (0.033)         &     (0.046)         &     (0.130)         &     (0.150)         \\
                ~~10 or more years&       0.041         &      -0.106\sym{**} &      -0.035         &       0.011         \\
                            &     (0.033)         &     (0.048)         &     (0.153)         &     (0.165)         \\ [6pt]
                Adjusted R\textsuperscript{2} & 0.008 & 0.001 & 0.012& 0.015 \\
                No. of observations & \multicolumn{1}{c}{\hspace{2mm}\num{38772}} & \multicolumn{1}{c}{\hspace{2mm}\num{44790}} & \multicolumn{1}{c}{\hspace{2mm}\num{38772}} & \multicolumn{1}{c}{\hspace{2mm}\num{44790}} \\
                Mean of dependent variable & 0.118 & 0.141 & 0.402 & 0.549 \\
				\bottomrule	
				\insertTableNotes
			\end{tabularx}
		\end{ThreePartTable}
	\end{table}
\end{singlespace}
\vfill
\clearpage


\vspace*{\fill}
\begin{singlespace}
	\begin{table}[!ht]
		\begin{ThreePartTable}
			\def\sym#1{\ifmmode^{#1}\else\(^{#1}\)\fi}
			\footnotesize
			\begin{TableNotes}[flushleft]\setlength\labelsep{0pt}\footnotesize\justifying
				\item\textit{Notes}: \sym{***} significant at 1\% level; \sym{**} significant at 5\% level; \sym{*} significant at 10\% level. All specifications include an intercept, year and region fixed effects, degree of urbanisation, education and marital and activity status. Heterokedasticity-robust standard errors in parentheses.
				\item\textit{Source}: Authors' analysis from national health surveys.				
			\end{TableNotes}
			\begin{tabularx}{\textwidth}{X *{4}{S[table-column-width=2cm]}}
				\caption{Age, immigrant arrival cohort and assimilation effects (OLS) in visits to GP and specialist (foreign-born population from Africa)} \label{Table A7}\\
				\toprule
				&\multicolumn{1}{c}{(I)}&\multicolumn{1}{c}{(II)}&\multicolumn{1}{c}{(III)}&\multicolumn{1}{c}{(IV)}\\[1ex]
				&\multicolumn{2}{c}{No. of visits to GP}&\multicolumn{2}{c}{No. of visits to specialist} \\[1ex]
				&\multicolumn{1}{c}{Men}&\multicolumn{1}{c}{Women}&\multicolumn{1}{c}{Men}&\multicolumn{1}{c}{Women}\\
				\midrule
                ~~Age       &       0.027\sym{***}&      -0.001         &       0.020\sym{***}&       0.004         \\
                            &     (0.006)         &     (0.006)         &     (0.006)         &     (0.003)         \\
                ~~$\text{Age}^2 /100$&      -0.043\sym{***}&       0.010         &      -0.032\sym{***}&      -0.003         \\
                            &     (0.012)         &     (0.013)         &     (0.011)         &     (0.007)         \\
                ~~$\text{Age}^3$ /10000&       0.026\sym{***}&      -0.005         &       0.016\sym{***}&      -0.000         \\
                            &     (0.008)         &     (0.008)         &     (0.006)         &     (0.004)         \\
                ~~$\text{Age} \times \text{migrant}$&      -0.038         &      -0.060         &       0.002         &       0.016         \\
                            &     (0.028)         &     (0.039)         &     (0.008)         &     (0.022)         \\
                ~~$\text{Age}/100 \times \text{migrant}$&       0.088         &       0.103         &      -0.010         &      -0.044         \\
                            &     (0.065)         &     (0.090)         &     (0.016)         &     (0.051)         \\
                ~~$\text{Age}/\num{10000} \times \text{migrant}$&      -0.062         &      -0.053         &       0.007         &       0.036         \\
                            &     (0.044)         &     (0.062)         &     (0.011)         &     (0.036)         \\
				Immigrant arrival cohort&&& \\ [6pt]
				~~Pre--1996 &       0.003         &       0.293         &      -0.038         &       0.107         \\
				&     (0.118)         &     (0.227)         &     (0.062)         &     (0.277)         \\
				~~1996--2007&       0.081         &       0.507\sym{**} &      -0.013         &       0.152         \\
				&     (0.110)         &     (0.210)         &     (0.063)         &     (0.282)         \\
				~~2008--2020&       0.100         &       0.272         &       0.005         &       0.066         \\
				&     (0.098)         &     (0.194)         &     (0.048)         &     (0.143)         \\
         		Time of residence in Spain&&& \\ [6pt]
                ~~5--9 years&      -0.012         &      -0.069         &      -0.025         &      -0.211         \\
                            &     (0.089)         &     (0.139)         &     (0.048)         &     (0.218)         \\
                ~~10 or more years&       0.047         &       0.080         &       0.025         &      -0.197         \\
                            &     (0.090)         &     (0.154)         &     (0.055)         &     (0.281)         \\ [6pt]
                Adjusted R\textsuperscript{2} & 0.052 & 0.032 & 0.031& 0.033 \\
                No. of observations & \multicolumn{1}{c}{\hspace{2mm}\num{40936}} & \multicolumn{1}{c}{\hspace{2mm}\num{46993}} & \multicolumn{1}{c}{\hspace{2mm}\num{40936}} & \multicolumn{1}{c}{\hspace{2mm}\num{46993}} \\
                Mean of dependent variable & 0.293 & 0.410 & 0.107 & 0.156 \\
				\bottomrule	
				\insertTableNotes
			\end{tabularx}
		\end{ThreePartTable}
	\end{table}
\end{singlespace}
\vfill
\clearpage


\vspace*{\fill}
\begin{singlespace}
	\begin{table}[!h]
		\begin{ThreePartTable}
			\def\sym#1{\ifmmode^{#1}\else\(^{#1}\)\fi}
			\footnotesize
			\begin{TableNotes}[flushleft]\setlength\labelsep{0pt}\footnotesize\justifying
				\item\textit{Notes}: \sym{***} significant at 1\% level; \sym{**} significant at 5\% level; \sym{*} significant at 10\% level. All specifications include an intercept, year and region fixed effects, degree of urbanisation, education and marital and activity status. Heterokedasticity-robust standard errors in parentheses.
				\item\textit{Source}: Authors' analysis from national health surveys.				
			\end{TableNotes}
			\begin{tabularx}{\textwidth}{X *{4}{S[table-column-width=2cm]}}
				\caption{Age, immigrant arrival cohort and assimilation effects (OLS) in hospital stays and visits to emergency care (foreign-born population from Africa)} \label{Table A8}\\
				\toprule
				&\multicolumn{1}{c}{(I)}&\multicolumn{1}{c}{(II)}&\multicolumn{1}{c}{(III)}&\multicolumn{1}{c}{(IV)}\\[1ex]
				&\multicolumn{2}{c}{No. of hospitalisations}&\multicolumn{2}{c}{No. of visits to emergency care} \\[1ex]
				&\multicolumn{1}{c}{Men}&\multicolumn{1}{c}{Women}&\multicolumn{1}{c}{Men}&\multicolumn{1}{c}{Women}\\
				\midrule
                ~~Age       &       0.039\sym{**} &       0.029         &       0.060\sym{***}&       0.030\sym{**} \\
                            &     (0.018)         &     (0.021)         &     (0.010)         &     (0.014)         \\
                ~~$\text{Age}^2 /100$&      -0.074\sym{**} &      -0.052         &      -0.134\sym{***}&      -0.084\sym{***}\\
                            &     (0.033)         &     (0.041)         &     (0.020)         &     (0.028)         \\
                ~~$\text{Age}^3$ /10000&       0.046\sym{**} &       0.030         &       0.088\sym{***}&       0.061\sym{***}\\
                            &     (0.019)         &     (0.023)         &     (0.012)         &     (0.016)         \\
                ~~$\text{Age} \times \text{migrant}$&       0.002         &       0.028         &      -0.082         &       0.126\sym{**} \\
                            &     (0.027)         &     (0.108)         &     (0.060)         &     (0.054)         \\
                ~~$\text{Age}/100 \times \text{migrant}$&      -0.014         &      -0.137         &       0.170         &      -0.273\sym{**} \\
                            &     (0.069)         &     (0.305)         &     (0.142)         &     (0.129)         \\
                ~~$\text{Age}/\num{10000} \times \text{migrant}$&       0.014         &       0.121         &      -0.110         &       0.188\sym{**} \\
                            &     (0.054)         &     (0.230)         &     (0.099)         &     (0.094)         \\
				Immigrant arrival cohort&&& \\ [6pt]
				~~Pre--1996 &       0.033         &       0.590         &       0.438         &      -0.417         \\
				&     (0.058)         &     (0.449)         &     (0.364)         &     (0.315)         \\
				~~1996--2007&       0.033         &       0.531         &       0.200         &      -0.113         \\
				&     (0.061)         &     (0.334)         &     (0.221)         &     (0.287)         \\
				~~2008--2020&      -0.031         &       0.265         &       0.246         &      -0.390         \\
				&     (0.055)         &     (0.282)         &     (0.230)         &     (0.244)         \\
				Time of residence in Spain&&& \\ [6pt]
                ~~5--9 years&       0.035         &      -0.145         &       0.127         &      -0.270         \\
                            &     (0.027)         &     (0.130)         &     (0.174)         &     (0.256)         \\
                ~~10 or more years&      -0.010         &       0.206         &       0.176         &      -0.298         \\
                            &     (0.042)         &     (0.418)         &     (0.173)         &     (0.271)         \\ [6pt]
                Adjusted R\textsuperscript{2} & 0.008 & 0.002 & 0.012& 0.016 \\
                No. of observations & \multicolumn{1}{c}{\hspace{2mm}\num{38245}} & \multicolumn{1}{c}{\hspace{2mm}\num{43327}} & \multicolumn{1}{c}{\hspace{2mm}\num{38245}} & \multicolumn{1}{c}{\hspace{2mm}\num{43327}} \\
                Mean of dependent variable & 0.118 & 0.147 & 0.404 & 0.544 \\
				\bottomrule	
				\insertTableNotes
			\end{tabularx}
		\end{ThreePartTable}
	\end{table}
\end{singlespace}
\vfill
\clearpage


\begin{figure}[!ht]
	\footnotesize
	\caption{Differences in health care use between 35-year-old migrants after 10 years in Spain by arrival cohort (foreign-born population from EU15 countries)}
	\centering 
	\includegraphics[width=0.8\textwidth]{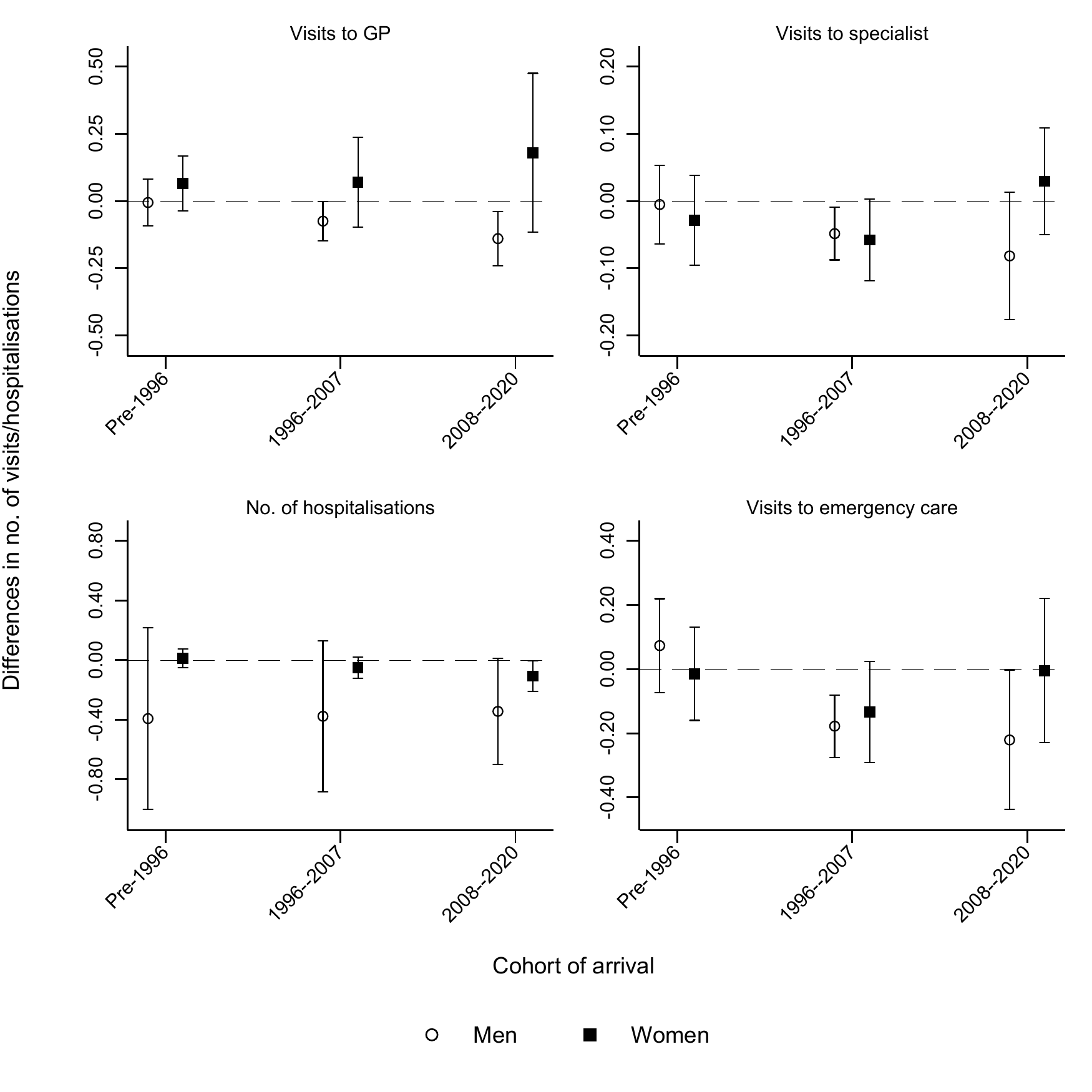} \\
	\justifying
	\noindent\textit{Note:} The graph shows point estimates and 90\%-level confidence intervals. We assume that migrants enter the country 10 years ago or earlier.\\
	\noindent\textit{Source}: Authors' analysis from results in Table~\ref{Table A1} and \ref{Table A2}.
	\label{Figure A1}
\end{figure}
\clearpage


\vspace*{\fill}
\begin{figure}[!ht]
	\footnotesize
	\caption{Differences in health care use between between 35-year-old migrants after 10 years in Spain by arrival cohort (foreign-born population from European countries other than EU15)}
	\centering 
	\includegraphics[width=0.8\textwidth]{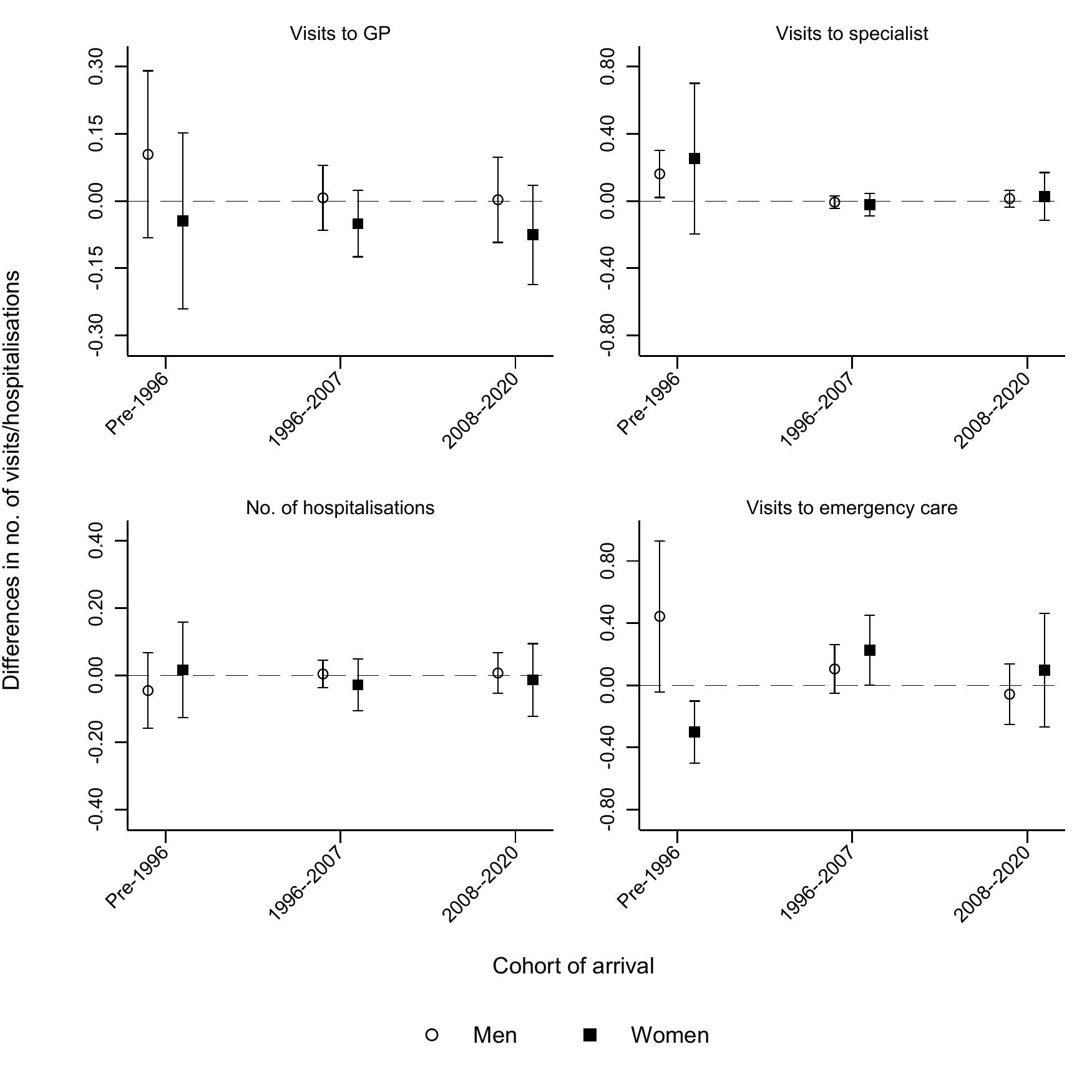} \\
	\justifying
	\noindent\textit{Note:} The graph shows point estimates and 90\%-level confidence intervals. We assume that migrants enter the country 10 years ago or earlier.\\
	\noindent\textit{Source}: Authors' analysis from results in Table~\ref{Table A3} and \ref{Table A4}.
	\label{Figure A2}
\end{figure}
\vfill
\clearpage


\vspace*{\fill}
\begin{figure}[!ht]
	\footnotesize
	\caption{Differences in health care use between 35-year-old migrants after 10 years in Spainby arrival cohort (foreign-born population from Latin America and the Caribbean)}
	\centering 
	\includegraphics[width=0.8\textwidth]{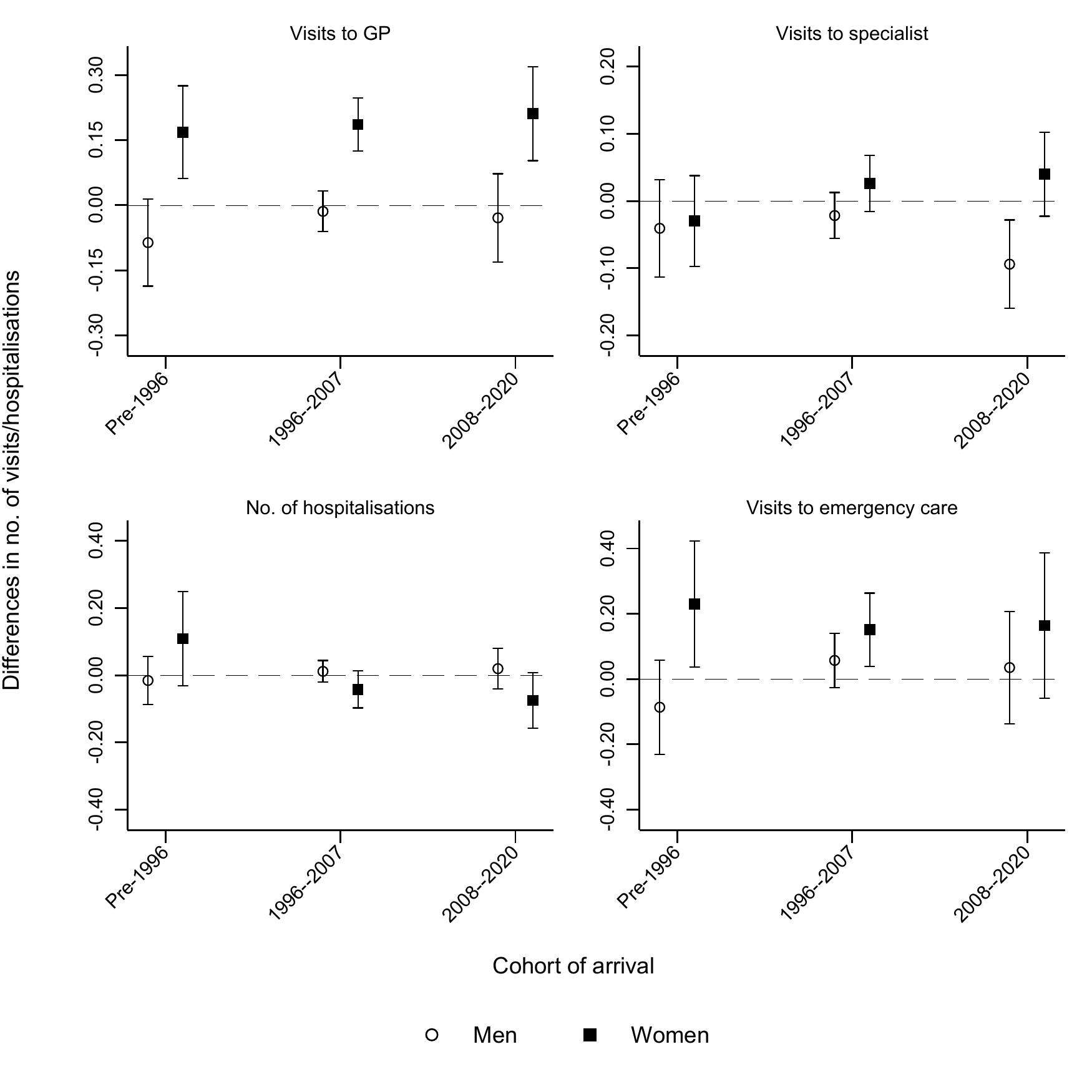} \\
	\justifying
	\noindent\textit{Note:} The graph shows point estimates and 90\%-level confidence intervals. We assume that migrants enter the country 10 years ago or earlier.\\
	\noindent\textit{Source}: Authors' analysis from results in Table~\ref{Table A5} and \ref{Table A6}.
	\label{Figure A3}
\end{figure}
\vfill
\clearpage


\vspace*{\fill}
\begin{figure}[!ht]
	\footnotesize
	\caption{Differences in health care use between 35-year-old migrants after 10 years in Spain by arrival cohort (foreign-born population from Africa)}
	\centering 
	\includegraphics[width=0.8\textwidth]{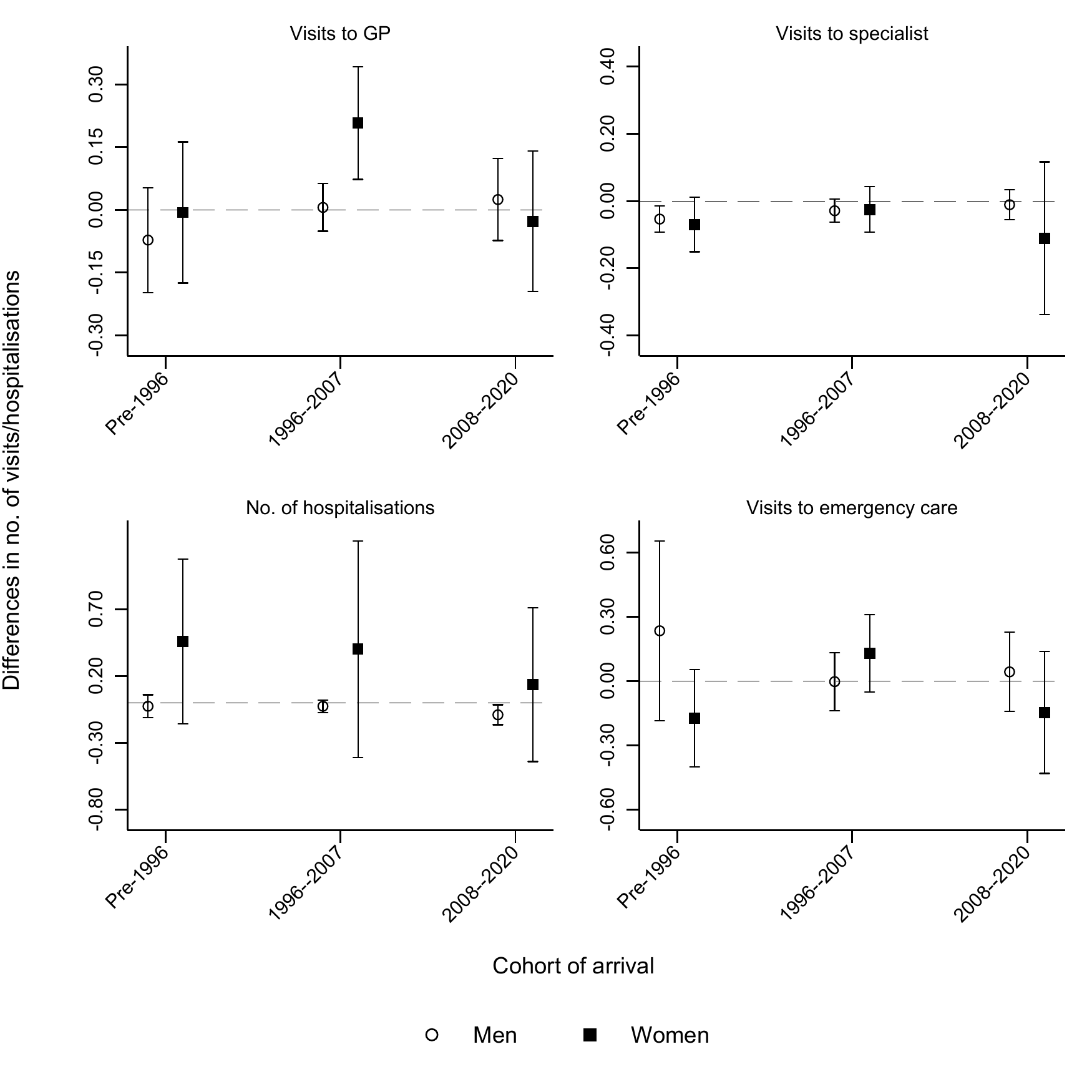} \\
	\justifying
	\noindent\textit{Note:} The graph shows point estimates and 90\%-level confidence intervals. We assume that migrants enter the country 10 years ago or earlier.\\
	\noindent\textit{Source}: Authors' analysis from results in Table~\ref{Table A5} and \ref{Table A6}.
	\label{Figure A4}
\end{figure}
\vfill
\clearpage



\vspace*{\fill}
\begin{singlespace}
	\begin{table}[!ht]
		\begin{ThreePartTable}
			\def\sym#1{\ifmmode^{#1}\else\(^{#1}\)\fi}
			\footnotesize
			\begin{TableNotes}[flushleft]\setlength\labelsep{0pt}\footnotesize\justifying
				\item\textit{Notes}: \sym{***} significant at 1\% level; \sym{**} significant at 5\% level; \sym{*} significant at 10\% level. All specifications include an intercept, year and region fixed effects, degree of urbanisation, education and marital and activity status. Heterokedasticity-robust standard errors in parentheses. R\textsuperscript{2} is the squared coefficient of correlation between the actual and the fitted values, as suggested by \textcite{zheng2000}.
				\item\textit{Source}: Authors' analysis from national health surveys.				
			\end{TableNotes}
			\begin{tabularx}{\textwidth}{X *{4}{S[table-column-width=2cm]}}
				\caption{Robustness checks (I): Immigrant arrival cohort and assimilation effects (Poisson-regression estimates) in visits to GP and specialist} \label{Table A9}\\
				\toprule
				&\multicolumn{1}{c}{(I)}&\multicolumn{1}{c}{(II)}&\multicolumn{1}{c}{(III)}&\multicolumn{1}{c}{(IV)}\\[1ex]
				&\multicolumn{2}{c}{No. of visits to GP}&\multicolumn{2}{c}{No. of visits to specialist} \\[1ex]
				&\multicolumn{1}{c}{Men}&\multicolumn{1}{c}{Women}&\multicolumn{1}{c}{Men}&\multicolumn{1}{c}{Women}\\
				\midrule
				Age effects and interactions&&& \\ [6pt]
                ~~Age       &       0.096\sym{***}&       0.003         &       0.205\sym{***}&       0.051\sym{**} \\
                            &     (0.021)         &     (0.015)         &     (0.050)         &     (0.023)         \\
                ~~$\text{Age}^2 /100$&      -0.133\sym{***}&       0.021         &      -0.342\sym{***}&      -0.062         \\
                            &     (0.041)         &     (0.029)         &     (0.097)         &     (0.044)         \\
                ~~$\text{Age}^3$ /10000&       0.068\sym{***}&      -0.014         &       0.175\sym{***}&       0.022         \\
                            &     (0.024)         &     (0.017)         &     (0.055)         &     (0.026)         \\
                ~~$\text{Age} \times \text{migrant}$&      -0.035         &       0.079\sym{**} &      -0.095         &       0.067         \\
                            &     (0.052)         &     (0.038)         &     (0.100)         &     (0.065)         \\
                ~~$\text{Age}/100 \times \text{migrant}$&       0.062         &      -0.169\sym{**} &       0.193         &      -0.142         \\
                            &     (0.112)         &     (0.082)         &     (0.211)         &     (0.137)         \\
                ~~$\text{Age}/\num{10000} \times \text{migrant}$&      -0.033         &       0.100\sym{*}  &      -0.124         &       0.090         \\
                            &     (0.074)         &     (0.055)         &     (0.139)         &     (0.090)         \\
				Immigrant arrival cohort&&& \\ [6pt]
				~~Pre--1996 &       0.246         &      -0.389\sym{*}  &       0.681         &      -0.480         \\
				&     (0.298)         &     (0.236)         &     (0.578)         &     (0.531)         \\
				~~1996--2007&       0.361         &      -0.251         &       0.609         &      -0.383         \\
				&     (0.278)         &     (0.210)         &     (0.540)         &     (0.485)         \\
				~~2008--2020&       0.129         &      -0.378\sym{*}  &      -0.340         &      -0.645\sym{*}  \\
				&     (0.244)         &     (0.197)         &     (0.498)         &     (0.349)         \\
				Time of residence in Spain&&& \\ [6pt]
                ~~5--9 years&      -0.271         &       0.075         &      -0.367         &      -0.271         \\
                            &     (0.203)         &     (0.149)         &     (0.394)         &     (0.350)         \\
                ~~10 or more years&      -0.184         &       0.282\sym{*}  &      -0.516         &       0.136         \\
                            &     (0.207)         &     (0.152)         &     (0.386)         &     (0.371)         \\ [6pt]
                R\textsuperscript{2} & 0.053 & 0.034 & 0.039& 0.035 \\
                No. of observations & \multicolumn{1}{c}{\hspace{2mm}\num{40936}} & \multicolumn{1}{c}{\hspace{2mm}\num{46993}} & \multicolumn{1}{c}{\hspace{2mm}\num{40936}} & \multicolumn{1}{c}{\hspace{2mm}\num{46993}} \\
                Mean of dependent variable & 0.293 & 0.410 & 0.107 & 0.156 \\
				\bottomrule	
				\insertTableNotes
			\end{tabularx}
		\end{ThreePartTable}
	\end{table}
\end{singlespace}
\vfill
\clearpage


\vspace*{\fill}
\begin{singlespace}
	\begin{table}[!ht]
		\begin{ThreePartTable}
			\def\sym#1{\ifmmode^{#1}\else\(^{#1}\)\fi}
			\footnotesize
			\begin{TableNotes}[flushleft]\setlength\labelsep{0pt}\footnotesize\justifying
				\item\textit{Notes}: \sym{***} significant at 1\% level; \sym{**} significant at 5\% level; \sym{*} significant at 10\% level. All specifications include an intercept, year and region fixed effects, degree of urbanisation, education and marital and activity status. Heterokedasticity-robust standard errors in parentheses. R\textsuperscript{2} is the squared coefficient of correlation between the actual and the fitted values, as suggested by \textcite{zheng2000}.
				\item\textit{Source}: Authors' analysis from national health surveys.				
			\end{TableNotes}
			\begin{tabularx}{\textwidth}{X *{4}{S[table-column-width=2cm]}}
				\caption{Robustness checks (I): Immigrant arrival cohort and assimilation effects (Poisson-regression estimates) in hospital stays and visits to emergency care} \label{Table A10}\\
				\toprule
				&\multicolumn{1}{c}{(I)}&\multicolumn{1}{c}{(II)}&\multicolumn{1}{c}{(III)}&\multicolumn{1}{c}{(IV)}\\[1ex]
				&\multicolumn{2}{c}{No. of hospitalisations}&\multicolumn{2}{c}{No. of visits to emergency care} \\[1ex]
				&\multicolumn{1}{c}{Men}&\multicolumn{1}{c}{Women}&\multicolumn{1}{c}{Men}&\multicolumn{1}{c}{Women}\\
				\midrule
				Age effects and interactions&&& \\ [6pt]
                ~~Age       &       0.319\sym{***}&       0.031         &       0.130\sym{***}&       0.060\sym{***}\\
                            &     (0.116)         &     (0.019)         &     (0.023)         &     (0.021)         \\
                ~~$\text{Age}^2 /100$&      -0.553\sym{**} &      -0.056         &      -0.287\sym{***}&      -0.161\sym{***}\\
                            &     (0.220)         &     (0.037)         &     (0.046)         &     (0.040)         \\
                ~~$\text{Age}^3$ /10000&       0.308\sym{**} &       0.032         &       0.187\sym{***}&       0.113\sym{***}\\
                            &     (0.123)         &     (0.021)         &     (0.026)         &     (0.024)         \\
                ~~$\text{Age} \times \text{migrant}$&      -0.214         &      -0.009         &      -0.047         &       0.102\sym{***}\\
                            &     (0.181)         &     (0.016)         &     (0.057)         &     (0.037)         \\
                ~~$\text{Age}/100 \times \text{migrant}$&       0.582         &       0.003         &       0.110         &      -0.208\sym{***}\\
                            &     (0.479)         &     (0.042)         &     (0.136)         &     (0.080)         \\
                ~~$\text{Age}/\num{10000} \times \text{migrant}$&      -0.436         &       0.004         &      -0.088         &       0.125\sym{**} \\
                            &     (0.360)         &     (0.031)         &     (0.098)         &     (0.053)         \\
				Immigrant arrival cohort&&& \\ [6pt]
				~~Pre--1996 &       0.890         &       0.243\sym{***}&       0.361         &      -0.521\sym{*}  \\
				&     (0.603)         &     (0.090)         &     (0.326)         &     (0.289)         \\
				~~1996--2007&       1.297\sym{**} &       0.170\sym{**} &       0.189         &      -0.316         \\
				&     (0.660)         &     (0.072)         &     (0.252)         &     (0.260)         \\
				~~2008--2020&       1.048         &       0.116\sym{**} &      -0.014         &      -0.399\sym{**} \\
				&     (0.755)         &     (0.053)         &     (0.207)         &     (0.202)         \\
				Time of residence in Spain&&& \\ [6pt]
                ~~5--9 years&      -1.015         &      -0.032         &      -0.006         &      -0.131         \\
                            &     (0.721)         &     (0.034)         &     (0.197)         &     (0.187)         \\
                ~~10 or more years&      -1.579\sym{**} &      -0.007         &       0.116         &       0.088         \\
                            &     (0.786)         &     (0.067)         &     (0.212)         &     (0.197)         \\ [6pt]
                R\textsuperscript{2} & 0.025 & 0.002 & 0.014& 0.015 \\
                No. of observations & \multicolumn{1}{c}{\hspace{2mm}\num{39944}} & \multicolumn{1}{c}{\hspace{2mm}\num{44820}} & \multicolumn{1}{c}{\hspace{2mm}\num{39944}} & \multicolumn{1}{c}{\hspace{2mm}\num{44820}} \\
                Mean of dependent variable & 0.115 & 0.134 & 0.403 & 0.550 \\
    			\bottomrule	
				\insertTableNotes
			\end{tabularx}
		\end{ThreePartTable}
	\end{table}
\end{singlespace}
\vfill
\clearpage



\vspace*{\fill}
\begin{singlespace}
	\begin{table}[!ht]
		\begin{ThreePartTable}
			\def\sym#1{\ifmmode^{#1}\else\(^{#1}\)\fi}
			\footnotesize
			\begin{TableNotes}[flushleft]\setlength\labelsep{0pt}\footnotesize\justifying
				\item\textit{Notes}: \sym{***} significant at 1\% level; \sym{**} significant at 5\% level; \sym{*} significant at 10\% level. All specifications include an intercept, year and region fixed effects, degree of urbanisation, education and marital and activity status. Heterokedasticity-robust standard errors in parentheses. 
				\item\textit{Source}: Authors' analysis from national health surveys.				
			\end{TableNotes}
			\begin{tabularx}{\textwidth}{X *{4}{S[table-column-width=2cm]}}
				\caption{Robustness checks (II): Immigrant arrival cohort and assimilation effects (OLS) in visits to GP and specialist (population with just NHS coverage)} \label{Table A11}\\
				\toprule
				&\multicolumn{1}{c}{(I)}&\multicolumn{1}{c}{(II)}&\multicolumn{1}{c}{(III)}&\multicolumn{1}{c}{(IV)}\\[1ex]
				&\multicolumn{2}{c}{No. of visits to GP}&\multicolumn{2}{c}{No. of visits to specialist} \\[1ex]
				&\multicolumn{1}{c}{Men}&\multicolumn{1}{c}{Women}&\multicolumn{1}{c}{Men}&\multicolumn{1}{c}{Women}\\
				\midrule
				Age effects and interactions&&& \\ [6pt]
                ~~Age       &       0.026\sym{***}&       0.001         &       0.019\sym{***}&       0.005         \\
                            &     (0.006)         &     (0.006)         &     (0.006)         &     (0.003)         \\
                ~~$\text{Age}^2 /100$&      -0.041\sym{***}&       0.008         &      -0.030\sym{***}&      -0.003         \\
                            &     (0.013)         &     (0.013)         &     (0.011)         &     (0.006)         \\
                ~~$\text{Age}^3$ /10000&       0.025\sym{***}&      -0.004         &       0.015\sym{**} &      -0.001         \\
                            &     (0.008)         &     (0.008)         &     (0.006)         &     (0.004)         \\
                ~~$\text{Age} \times \text{migrant}$&      -0.008         &       0.028\sym{**} &      -0.005         &       0.011         \\
                            &     (0.013)         &     (0.014)         &     (0.008)         &     (0.008)         \\
                ~~$\text{Age}/100 \times \text{migrant}$&       0.017         &      -0.060\sym{*}  &       0.008         &      -0.025         \\
                            &     (0.031)         &     (0.032)         &     (0.017)         &     (0.017)         \\
                ~~$\text{Age}/\num{10000} \times \text{migrant}$&      -0.012         &       0.035         &      -0.004         &       0.016         \\
                            &     (0.022)         &     (0.023)         &     (0.012)         &     (0.012)         \\
                Immigrant arrival cohort&&& \\ [6pt]
				~~Pre--1996 &       0.023         &      -0.119         &       0.027         &      -0.086         \\
				&     (0.063)         &     (0.078)         &     (0.041)         &     (0.064)         \\
				~~1996--2007&       0.064         &      -0.060         &       0.033         &      -0.076         \\
				&     (0.057)         &     (0.065)         &     (0.037)         &     (0.054)         \\
				~~2008--2020&       0.057         &      -0.096         &      -0.001         &      -0.077\sym{**} \\
				&     (0.050)         &     (0.059)         &     (0.032)         &     (0.034)         \\
				Time of residence in Spain&&& \\ [6pt]
                ~~5--9 years&      -0.054         &      -0.001         &      -0.013         &      -0.016         \\
                            &     (0.043)         &     (0.046)         &     (0.026)         &     (0.038)         \\
                ~~10 or more years&      -0.033         &       0.080\sym{*}  &      -0.012         &       0.041         \\
                            &     (0.046)         &     (0.048)         &     (0.030)         &     (0.048)         \\ [6pt]
                Adjusted R\textsuperscript{2} & 0.054 & 0.032 & 0.032& 0.033 \\
                No. of observations & \multicolumn{1}{c}{\hspace{2mm}\num{36497}} & \multicolumn{1}{c}{\hspace{2mm}\num{43294}} & \multicolumn{1}{c}{\hspace{2mm}\num{36497}} & \multicolumn{1}{c}{\hspace{2mm}\num{43294}} \\
                Mean of dependent variable & 0.298 & 0.419 & 0.104 & 0.151 \\
				\bottomrule	
				\insertTableNotes
			\end{tabularx}
		\end{ThreePartTable}
	\end{table}
\end{singlespace}
\vfill
\clearpage


\vspace*{\fill}
\begin{singlespace}
	\begin{table}[!ht]
		\begin{ThreePartTable}
			\def\sym#1{\ifmmode^{#1}\else\(^{#1}\)\fi}
			\footnotesize
			\begin{TableNotes}[flushleft]\setlength\labelsep{0pt}\footnotesize\justifying
				\item\textit{Notes}: \sym{***} significant at 1\% level; \sym{**} significant at 5\% level; \sym{*} significant at 10\% level. All specifications include an intercept, year and region fixed effects, degree of urbanisation, education and marital and activity status. Heterokedasticity-robust standard errors in parentheses.
				\item\textit{Source}: Authors' analysis from national health surveys.				
			\end{TableNotes}
			\begin{tabularx}{\textwidth}{X *{4}{S[table-column-width=2cm]}}
				\caption{Robustness checks (II): Immigrant arrival cohort and assimilation effects (Poisson-regression estimates) in hospital stays and visits to emergency care (population with just NHS coverage)} \label{Table A12}\\
				\toprule
				&\multicolumn{1}{c}{(I)}&\multicolumn{1}{c}{(II)}&\multicolumn{1}{c}{(III)}&\multicolumn{1}{c}{(IV)}\\[1ex]
				&\multicolumn{2}{c}{No. of hospitalisations}&\multicolumn{2}{c}{No. of visits to emergency care} \\[1ex]
				&\multicolumn{1}{c}{Men}&\multicolumn{1}{c}{Women}&\multicolumn{1}{c}{Men}&\multicolumn{1}{c}{Women}\\
				\midrule
				Age effects and interactions&&& \\ [6pt]
                ~~Age       &       0.046\sym{**} &       0.030         &       0.059\sym{***}&       0.032\sym{**} \\
                            &     (0.019)         &     (0.021)         &     (0.011)         &     (0.015)         \\
                ~~$\text{Age}^2 /100$&      -0.088\sym{**} &      -0.055         &      -0.130\sym{***}&      -0.088\sym{***}\\
                            &     (0.036)         &     (0.040)         &     (0.021)         &     (0.028)         \\
                ~~$\text{Age}^3$ /10000&       0.053\sym{***}&       0.031         &       0.085\sym{***}&       0.063\sym{***}\\
                            &     (0.020)         &     (0.023)         &     (0.012)         &     (0.017)         \\
                ~~$\text{Age} \times \text{migrant}$&      -0.045         &      -0.008         &      -0.026         &       0.059\sym{**} \\
                            &     (0.041)         &     (0.018)         &     (0.026)         &     (0.024)         \\
                ~~$\text{Age}/100 \times \text{migrant}$&       0.119         &       0.000         &       0.057         &      -0.123\sym{**} \\
                            &     (0.107)         &     (0.047)         &     (0.059)         &     (0.052)         \\
                ~~$\text{Age}/\num{10000} \times \text{migrant}$&      -0.088         &       0.006         &      -0.044         &       0.078\sym{**} \\
                            &     (0.077)         &     (0.035)         &     (0.041)         &     (0.035)         \\	
				Immigrant arrival cohort&&& \\ [6pt]
				~~Pre--1996 &       0.207         &       0.254\sym{***}&       0.172         &      -0.266         \\
				&     (0.209)         &     (0.097)         &     (0.153)         &     (0.176)         \\
				~~1996--2007&       0.275         &       0.173\sym{**} &       0.110         &      -0.160         \\
				&     (0.266)         &     (0.075)         &     (0.116)         &     (0.162)         \\
				~~2008--2020&       0.270         &       0.112\sym{**} &       0.035         &      -0.217\sym{*}  \\
				&     (0.300)         &     (0.057)         &     (0.098)         &     (0.132)         \\
             	Time of residence in Spain&&& \\ [6pt]
                ~~5--9 years&      -0.199         &      -0.032         &       0.003         &      -0.090         \\
                            &     (0.255)         &     (0.037)         &     (0.076)         &     (0.111)         \\
                ~~10 or more years&      -0.280         &      -0.008         &       0.058         &       0.025         \\
                            &     (0.276)         &     (0.072)         &     (0.085)         &     (0.116)         \\ [6pt]
                Adjusted R\textsuperscript{2} & 0.007 & 0.001 & 0.012& 0.014 \\
                No. of observations & \multicolumn{1}{c}{\hspace{2mm}\num{36497}} & \multicolumn{1}{c}{\hspace{2mm}\num{43294}} & \multicolumn{1}{c}{\hspace{2mm}\num{36497}} & \multicolumn{1}{c}{\hspace{2mm}\num{43294}} \\
                Mean of dependent variable & 0.120 & 0.145 & 0.411 & 0.559 \\
    			\bottomrule	
				\insertTableNotes
			\end{tabularx}
		\end{ThreePartTable}
	\end{table}
\end{singlespace}
\vfill
\clearpage



\vspace*{\fill}
\begin{singlespace}
	\begin{table}[!ht]
		\begin{ThreePartTable}
			\def\sym#1{\ifmmode^{#1}\else\(^{#1}\)\fi}
			\footnotesize
			\begin{TableNotes}[flushleft]\setlength\labelsep{0pt}\footnotesize\justifying
				\item\textit{Notes}: \sym{***} significant at 1\% level; \sym{**} significant at 5\% level; \sym{*} significant at 10\% level. All specifications include an intercept, year and region fixed effects, degree of urbanisation, education and marital and activity status. Heterokedasticity-robust standard errors in parentheses. 
				\item\textit{Source}: Authors' analysis from national health surveys.				
			\end{TableNotes}
			\begin{tabularx}{\textwidth}{X *{4}{S[table-column-width=2cm]}}
				\caption{Robustness checks (III): Immigrant arrival cohort and assimilation effects (OLS) in visits to GP and specialist (population aged less than 64 years old)} \label{Table A13}\\
				\toprule
				&\multicolumn{1}{c}{(I)}&\multicolumn{1}{c}{(II)}&\multicolumn{1}{c}{(III)}&\multicolumn{1}{c}{(IV)}\\[1ex]
				&\multicolumn{2}{c}{No. of visits to GP}&\multicolumn{2}{c}{No. of visits to specialist} \\[1ex]
				&\multicolumn{1}{c}{Men}&\multicolumn{1}{c}{Women}&\multicolumn{1}{c}{Men}&\multicolumn{1}{c}{Women}\\
				\midrule
				Age effects and interactions&&& \\ [6pt]
                ~~Age       &       0.029\sym{***}&       0.046\sym{***}&       0.011         &       0.016\sym{**} \\
                            &     (0.008)         &     (0.012)         &     (0.007)         &     (0.008)         \\
                ~~$\text{Age}^2 /100$&      -0.048\sym{**} &      -0.114\sym{***}&      -0.005         &      -0.034         \\
                            &     (0.022)         &     (0.031)         &     (0.017)         &     (0.022)         \\
                ~~$\text{Age}^3$ /10000&       0.031\sym{*}  &       0.097\sym{***}&      -0.008         &       0.024         \\
                            &     (0.018)         &     (0.026)         &     (0.014)         &     (0.018)         \\
                ~~$\text{Age} \times \text{migrant}$&       0.023         &      -0.009         &       0.007         &      -0.016         \\
                            &     (0.022)         &     (0.029)         &     (0.015)         &     (0.017)         \\
                ~~$\text{Age}/100 \times \text{migrant}$&      -0.071         &       0.038         &      -0.024         &       0.051         \\
                            &     (0.063)         &     (0.080)         &     (0.042)         &     (0.048)         \\
                ~~$\text{Age}/\num{10000} \times \text{migrant}$&       0.064         &      -0.043         &       0.022         &      -0.048         \\
                            &     (0.056)         &     (0.069)         &     (0.036)         &     (0.041)         \\
				Immigrant arrival cohort&&& \\ [6pt]
				~~Pre--1996 &       0.017         &      -0.103         &       0.022         &      -0.035         \\
				&     (0.065)         &     (0.078)         &     (0.041)         &     (0.064)         \\
				~~1996--2007&       0.021         &      -0.018         &       0.011         &      -0.037         \\
				&     (0.056)         &     (0.066)         &     (0.036)         &     (0.052)         \\
				~~2008--2020&      -0.010         &      -0.069         &      -0.022         &      -0.035         \\
				&     (0.049)         &     (0.058)         &     (0.033)         &     (0.034)         \\
				Time of residence in Spain&&& \\ [6pt]
                ~~5--9 years&      -0.041         &       0.016         &      -0.012         &      -0.019         \\
                            &     (0.038)         &     (0.044)         &     (0.025)         &     (0.037)         \\
                ~~10 or more years&      -0.022         &       0.081\sym{*}  &      -0.011         &       0.039         \\
                            &     (0.041)         &     (0.047)         &     (0.028)         &     (0.046)         \\ [6pt]
                Adjusted R\textsuperscript{2} & 0.023 & 0.020 & 0.024& 0.033 \\
                No. of observations & \multicolumn{1}{c}{\hspace{2mm}\num{30295}} & \multicolumn{1}{c}{\hspace{2mm}\num{31240}} & \multicolumn{1}{c}{\hspace{2mm}\num{30295}} & \multicolumn{1}{c}{\hspace{2mm}\num{31240}} \\
                Mean of dependent variable & 0.236 & 0.357 & 0.092 & 0.149 \\
				\bottomrule	
				\insertTableNotes
			\end{tabularx}
		\end{ThreePartTable}
	\end{table}
\end{singlespace}
\vfill
\clearpage


\vspace*{\fill}
\begin{singlespace}
	\begin{table}[!ht]
		\begin{ThreePartTable}
			\def\sym#1{\ifmmode^{#1}\else\(^{#1}\)\fi}
			\footnotesize
			\begin{TableNotes}[flushleft]\setlength\labelsep{0pt}\footnotesize\justifying
				\item\textit{Notes}: \sym{***} significant at 1\% level; \sym{**} significant at 5\% level; \sym{*} significant at 10\% level. All specifications include an intercept, year and region fixed effects, degree of urbanisation, education and marital and activity status. Heterokedasticity-robust standard errors in parentheses.
				\item\textit{Source}: Authors' analysis from national health surveys.				
			\end{TableNotes}
			\begin{tabularx}{\textwidth}{X *{4}{S[table-column-width=2cm]}}
				\caption{Robustness checks (III): Immigrant arrival cohort and assimilation effects (Poisson-regression estimates) in hospital stays and visits to emergency care (population aged less than 64 years old)} \label{Table A14}\\
				\toprule
				&\multicolumn{1}{c}{(I)}&\multicolumn{1}{c}{(II)}&\multicolumn{1}{c}{(III)}&\multicolumn{1}{c}{(IV)}\\[1ex]
				&\multicolumn{2}{c}{No. of hospitalisations}&\multicolumn{2}{c}{No. of visits to emergency care} \\[1ex]
				&\multicolumn{1}{c}{Men}&\multicolumn{1}{c}{Women}&\multicolumn{1}{c}{Men}&\multicolumn{1}{c}{Women}\\
				\midrule
				Age effects and interactions&&& \\ [6pt]
                ~~Age       &       0.045\sym{***}&       0.030         &       0.053\sym{***}&       0.105\sym{***}\\
                            &     (0.014)         &     (0.040)         &     (0.015)         &     (0.027)         \\
                ~~$\text{Age}^2 /100$&      -0.082\sym{***}&      -0.058         &      -0.110\sym{***}&      -0.280\sym{***}\\
                            &     (0.027)         &     (0.108)         &     (0.039)         &     (0.067)         \\
                ~~$\text{Age}^3$ /10000&       0.049\sym{**} &       0.039         &       0.066\sym{**} &       0.220\sym{***}\\
                            &     (0.020)         &     (0.092)         &     (0.033)         &     (0.052)         \\
                ~~$\text{Age} \times \text{migrant}$&       0.063         &       0.075         &       0.003         &       0.046         \\
                            &     (0.050)         &     (0.076)         &     (0.046)         &     (0.052)         \\
                ~~$\text{Age}/100 \times \text{migrant}$&      -0.197         &      -0.233         &      -0.033         &      -0.092         \\
                            &     (0.161)         &     (0.220)         &     (0.124)         &     (0.145)         \\
                ~~$\text{Age}/\num{10000} \times \text{migrant}$&       0.198         &       0.207         &       0.042         &       0.056         \\
                            &     (0.167)         &     (0.191)         &     (0.103)         &     (0.125)         \\
				Immigrant arrival cohort&&& \\ [6pt]
				~~Pre--1996 &       0.065         &      -0.222\sym{**} &       0.016         &      -0.184\sym{***}\\
				&     (0.104)         &     (0.110)         &     (0.070)         &     (0.065)         \\
				~~1996--2007&       0.137         &      -0.204\sym{**} &       0.035         &      -0.128\sym{***}\\
				&     (0.093)         &     (0.089)         &     (0.057)         &     (0.048)         \\
				~~2008--2020&       0.122\sym{*}  &      -0.179\sym{**} &      -0.037         &      -0.114\sym{***}\\
				&     (0.071)         &     (0.077)         &     (0.054)         &     (0.039)         \\
				Time of residence in Spain&&& \\ [6pt]
                ~~5--9 years&      -0.183         &      -0.036         &      -0.008         &      -0.086         \\
                            &     (0.225)         &     (0.035)         &     (0.072)         &     (0.105)         \\
                ~~10 or more years&      -0.227         &      -0.002         &       0.063         &       0.053         \\
                            &     (0.243)         &     (0.078)         &     (0.081)         &     (0.111)         \\ [6pt]
                Adjusted R\textsuperscript{2} & 0.005 & 0.001 & 0.016& 0.014 \\
                No. of observations & \multicolumn{1}{c}{\hspace{2mm}\num{30295}} & \multicolumn{1}{c}{\hspace{2mm}\num{31240}} & \multicolumn{1}{c}{\hspace{2mm}\num{30295}} & \multicolumn{1}{c}{\hspace{2mm}\num{31240}} \\
                Mean of dependent variable & 0.087 & 0.129 & 0.390 & 0.556 \\
				\bottomrule	
				\insertTableNotes
			\end{tabularx}
		\end{ThreePartTable}
	\end{table}
\end{singlespace}
\vfill
	
\end{document}